\documentclass[11pt]{article}
\usepackage{graphicx, epsfig, amssymb}% Include figure files
\usepackage{amsfonts,amssymb,amsmath} % amsmath allows to use \eqref{}
\usepackage{latexsym}
\usepackage[english]{babel}

\newcommand{\be}{\begin{equation}}
\newcommand{\ee}{\end{equation}}

\newcommand{\beq}{\begin{eqnarray}}
\newcommand{\eeq}{\end{eqnarray}}

\newcommand{\T}{{\cal T}}

\newcommand{\eg}{{\it e.g.,}\ }
\newcommand{\ie}{{\it i.e.,}\ }
\newcommand{\p}{\partial}

\newcommand{\lp}{\left(}
\newcommand{\rp}{\right)}
\newcommand{\lpp}{\left[}
\newcommand{\rpp}{\right]}

\font\mybb=msbm10 at 12pt
\def\bb#1{\hbox{\mybb#1}}
\def\bZ {\bb{Z}}

\newcommand{\tE}{\widetilde{E}}

\newcommand{\tJ}{\widetilde{J}}
\newcommand{\tS}{\widetilde{S}}
\newcommand{\tT}{\widetilde{T}}
\newcommand{\tO}{\widetilde{\Omega}}

\newcommand{\tloc}{\mathcal{T}}

\newcommand{\ri}{R_\mathrm{i}}
\newcommand{\ro}{R_\mathrm{o}}
\newcommand{\vi}{v_\mathrm{i}}
\newcommand{\vo}{v_\mathrm{o}}
\newcommand{\z}{\tilde{r}}

\newcommand{\lrp}[1]{\left ( #1 \right )}

\newcommand{\lrpp}[1]{\left [ #1 \right ]}

\begin{document}

\setlength{\unitlength}{1mm}

\thispagestyle{empty}
%\rightline{\small hep-th/08}
 \vspace*{0.5cm}

\begin{center}

{\bf \Large Phase diagram for non-axisymmetric plasma balls}

 \vspace*{2cm}

{\bf Vitor Cardoso,}$^{1\,,\,2}\,$  {\bf \'Oscar
J.~C.~Dias,}$^{3}\,$ {\bf Jorge V. Rocha,}$^{1}$

\vspace*{0.5cm}

{\it $^1\,$ CENTRA, Dept. de F\'{\i}sica, Instituto
Superior T\'ecnico, \\
Av. Rovisco Pais 1, 1049-001 Lisboa, Portugal}\\[.3em]
{\it $^2\,$Dept. of Physics and Astronomy, The University of
Mississippi, \\
University, MS
38677-1848, USA}\\[.3em]
{\it $^3\,$DAMTP, Centre for Mathematical Sciences, University of
Cambridge,\\
Wilberforce Road, Cambridge CB3 0WA,
United Kingdom}\\[.3em]

\vspace*{0.3cm} {\tt vitor.cardoso@ist.utl.pt,
O.Dias@damtp.cam.ac.uk, jorge.v.rocha@ist.utl.pt}

\vspace*{2cm}

 {\bf ABSTRACT}
\end{center}
\vspace*{0.5cm}

Plasma balls and rings emerge as fluid holographic duals of black
holes and black rings in the hydrodynamic/gravity correspondence for
the Scherk-Schwarz AdS system.
Recently, plasma balls spinning above a critical rotation were found
to be unstable against $m$-lobed perturbations.
In the phase diagram of stationary solutions the threshold of the
instability signals a bifurcation to a new phase of non-axisymmetric
configurations.
We find explicitly this family of solutions and represent them in the
phase diagram.
We discuss the implications of our results for the gravitational system.
Rotating non-axisymmetric black holes necessarily radiate gravitational
waves.
We thus emphasize that it would be important, albeit possibly out of present
reach, to have a better understanding of the hydrodynamic description of
gravitational waves and of the gravitational interaction between two bodies.
We also argue that it might well be that a non-axisymmetric $m$-lobed
instability is also present in Myers-Perry black holes for rotations below
the recently found ultraspinning instability.

\noindent

%\keywords{AdS-CFT correspondence, black holes}

\vfill \setcounter{page}{0} \setcounter{footnote}{0}
\newpage

\tableofcontents

%\newpage
%%%%%%%%%%%%%%%%%%%%%%%%%%%%%%%%%%%%%%%%%%%%%%%%%%%%%%%%%%%%%%%%%%%%%%%%%%%%%%%%%%%%%%%%%%%%%%%%%%%%%%%%%%%%%%%%%%%%%%%%%%%%%%%%%%%%%%%%%%%%%%%
\setcounter{equation}{0}
\section{Introduction}
%%%%%%%%%%%%%%%%%%%%%%%%%%%%%%%%%%%%%%%%%%%%%%%%%%%%%%%%%%%%%%%%%%%%%%%%%%%%%%%%%%%%%%%%%%%%%%%%%%%%%%%%%%%%%%%%%%%%%%%%%%%%%%%%%%%%%%%%%%%%%%%

The idea of an hypothetical connection between black holes and fluid
balls has been the subject of several studies along the decades.
Nevertheless, for a long time no formal map between these two
distinct systems was available. (Although the membrane paradigm,
whereby a black hole horizon is mimicked by a stretched fluid
membrane, provided an interesting analogue model with useful
applications in astrophysical systems). However, the state of the
art changed radically in recent years. Motivated by the AdS/CFT
duality and by Landau's observation that any quantum field theory
should have an effective hydrodynamic description at high energy
densities (where the mean free path of the theory is small), a
rigorous duality between gravity and hydrodynamics was finally
found. Emerging with the path-breaking studies
\cite{Bhattacharyya:2008jc,Bhattacharyya:2008xc}, themselves
motivated by the ideas of \cite{kovtun}-\cite{Bhattacharyya:2007vs},
this duality has been since then successfully explored in a series
of works \cite{VanRaamsdonk:2008fp}-\cite{Figueras:2009iu}; for a
nice review and references see \cite{Rangamani:2009xk}.

The main idea is as follows
\cite{Bhattacharyya:2008jc,Bhattacharyya:2008xc,Rangamani:2009xk}.
Start with a black hole solution of Einstein-AdS gravity and divide
its horizon in several patches. Consider then the approximation
where in the tubewise region that goes from each horizon patch up to
the AdS boundary the geometry is approximately described by the line
element of a boosted planar black brane. This is parametrized by the
temperature $T$ and boost parameters $\beta_i$ along the timelike
and boundary spacelike coordinates $x^i$. When moving along the
several patches one allows the temperature and boosts to vary slowly
with the boundary coordinates. Obviously such a patched geometry
will generically not be a solution of Einstein-AdS gravity. However,
it will be a solution at any order of a perturbative expansion of
the Einstein-AdS field equations as long as at each order $n$ the
holographic stress tensor $T^{\mu\nu}_{(n)}$ satisfies certain
equations that turn out to be simply $\nabla_\nu
T^{\mu\nu}_{(n)}=0$, \ie it must be conserved (the expansion is done
assuming that the lengthscales of the temperature and boost
variations are much larger than the thermal lengthscale of the
system). This stress tensor is found by computing the extrinsic
curvature of a constant radial surface in the limit where this
surface approaches the AdS boundary. The idea that the stress tensor
must be conserved is certainly not new and was realized in previous
studies of holographic descriptions of gravity as a quantum field
theory on the boundary (namely in the AdS/CFT duality). However,
with the exception of a few explicit examples, this stress tensor
was {\it a priori} an arbitrary symmetric tensor in $d$ dimensions
with $d(d+1)/2$ components (eventually $d(d+1)/2-1$ if the quantum
field theory on the boundary is conformal and thus traceless). These
are too many degrees of freedom for the conservation equations to be
closed. What is new with the boosted black brane patching and the
perturbative analysis that follows is that in the process the form
of the stress tensor is highly and uniquely constrained. For
example, at leading order in the expansion it must have a perfect
fluid form. This reduces strongly the degrees of freedom and the
conservation equations now constitute a closed system: we have as
many equations as unknowns. Moreover, there is also a one-to-one map
between the bulk gravity solution and the boundary system. Working
out the conservation equations we further find that they boil down
to the equations of fluid dynamics.

These ideas were first developed to get a hydrodynamic description
of AdS/CFT duality. However they naturally extend to other
gravity/gauge theory dualities. Here we will be interested in dual
systems that have a confinement/deconfinement phase transition
\cite{Aharony:2005bm,Lahiri:2007ae}. A particular example, that is
the simplest one with these properties, is the Scherk-Schwarz (SS)
gravity/gauge system \cite{Witten:1998zw,Mateos:2007ay}. The reason
to study such a system is (at least) two-folded. First, from a
quantum field theory perspective, the SS gauge theory is
non-supersymmetric, non-conformal and has the above mentioned phase
transition. These are important properties that it shares with the
real world QCD. Therefore the SS system can be a useful toy model to
infer properties of QCD. In particular, having the gravity dual of
the SS gauge theory we can use the fact that this is a weak/strong
coupling duality to use the weak field gravity results to learn
about the strong field (non-perturbative) regime of the field
theory. Second, from a gravitational perspective, previous studies
of the hydrodynamic SS system revealed an interesting but {\it a
priori} unexpected feature
\cite{Aharony:2005bm}-\cite{Bhattacharya:2009gm}\cite{Caldarelli:2008ze}.
SS AdS gravity differs from the globally AdS gravity (some of the
main differences will be discussed in Section \ref{sec:Discussion}).
However, results from SS hydrodynamics are telling us that black
holes in this background have surprising similarities with globally
AdS black holes or even asymptotically flat black holes. We can
therefore use the SS hydrodynamic results to learn about these
latter black holes in more conventional backgrounds.

Concretely, in the long wavelength regime, the SS compactification
of a $(d+1)$-dimensional CFT has an effective description as a
$d$-dimensional fluid dynamics with an equation of state that
describes the SS plasma.  This system is dual to the SS
compactification of AdS$_{d+2}$ gravity. In short, the SS theory has
both a confined and a deconfined phase. In the gravity side of the
duality the deconfined phase corresponds to a black hole solution
localized in the infrared region of the holographic direction, while
the confined phase maps to the AdS soliton solution. They compete
with each other for the minimal free energy. In the neighborhood of
the confinement temperature the two phases co-exist in different
regions of the spacetime and are separated by a domain wall whose
tension provides the Israel junction conditions between the two
solutions. In the holographic boundary where the fluid lives, the
deconfined phase is described by a plasma lump immersed in the
confined sea phase. The plasma lump has a surface tension at its
boundary that is in correspondence with the domain wall tension in
the bulk. For a more detailed description of this system we refer
the reader to \cite{Lahiri:2007ae,Cardoso:2009bv}.

For the present discussion, we should keep in mind that plasma lumps
in $d$-dimensions correspond to SS AdS$_{d+2}$ black objects. For
example, axisymmetric plasma balls and plasma rings correspond,
respectively, to rotating black holes and black rings in SS
AdS$_{d+2}$ \cite{Lahiri:2007ae}. In a previous study, rotating
plasma balls were found to be unstable against $m$-lobed
perturbations~\cite{Cardoso:2009bv}. Starting with a static plasma
ball and increasing its rotation, the axisymmetric rotating plasma
ball becomes unstable first to a $2$-lobed (``peanut-like")
perturbation and then to $m$-lobed perturbations with $m>2$. This
conclusion followed from analyzing perturbations of the hydrodynamic
equations. Usually, such an instability signals a bifurcation point
to a new branch of stationary solutions (that obeys the symmetries
of the instability) in the phase diagram of stationary solutions.
Here, we will find that this is indeed the case and we will
construct explicitly the branch of $2$-lobed plasma balls in the
phase diagram of solutions.

The paper is organized as follows. Section~\ref{sec:Hydrodynamics}
reviews the relativistic hydrodynamics of $(2+1)$-dimensional, stationary
plasma configurations and displays the foundational equations for the
remainder of the article. The main new results are contained in
Section~\ref{sec:ballsRingsLobed}, where the non-axisymmetric profile
of the $m$-lobed plasmas is found and the phase diagram for stationary
plasmas is determined. For completeness, we briefly review the previuosly
known results for axisymmetric solutions, both plasma balls and rings. In
addition, we present a simple derivation that precludes the existence of
$m$-lobed plasma rings as perturbations of axisymmetric plasma rings. We
conclude in Section~\ref{sec:Discussion} with a discussion and some speculations
on the gravitational duals of these plasma lumps. Appendix~\ref{sec:NRpeanuts}
contains an analysis of the non-relativistic limit.
%%%%%%%%%%%%%%%%%%%%%%%%%%%%%%%%%%%%%%%%%%%%%%%%%%%%%%%%%%%%%%%%%%%%%%%%%%%%%%%%%%%%%%%%%%%%%%%%%%%%%%%%%%%%%%
\setcounter{equation}{0}
\section{Hydrodynamics of stationary plasma lumps \label{sec:Hydrodynamics}}
%%%%%%%%%%%%%%%%%%%%%%%%%%%%%%%%%%%%%%%%%%%%%%%%%%%%%%%%%%%%%%%%%%%%%%%%%%%%%%%%%%%%%%%%%%%%%%%%%%%%%%%%%%%%%%

In this section we review the relativistic hydrodynamic
equations that govern a Scherk-Schwarz plasma in a $3d$ Minkowski
background. We will be interested in plasma configurations in
mechanical and thermodynamic equilibrium. This includes
non-axisymmetric configurations. We discuss the conserved charges
for stationary plasmas that allow to represent the branches of
stationary plasmas in a phase diagram. We follow closely
\cite{Caldarelli:2008mv,Lahiri:2007ae}.

%%%%%%%%%%%%%%%%%%%%%%%%%%%%%%%%%%%%%%%%%%%%%%%%%%%%%%%%%%%%%%%%%%%%%%
\subsection{Relativistic hydrodynamic equations \label{sec:HydroEqs}}
%%%%%%%%%%%%%%%%%%%%%%%%%%%%%%%%%%%%%%%%%%%%%%%%%%%%%%%%%%%%%%%%%%%%%%

SS hydrodynamics is an effective description at long distances of SS
gauge theory, valid when the fluid variables vary on distance scales
that are large compared to the mean free path of the theory,
$l_\mathrm{mfp}\sim 1/T_c$. Expanding the holographic stress tensor
$T_{\mu\nu}$ in powers of derivatives of $u^{\mu}$, one finds that
at leading order $T_{\mu\nu}$ has a perfect fluid plus a surface
tension boundary contributions, while the dissipation effects due to
shear and bulk viscosity plus heat diffusion appear in the
next-to-leading order term.

The hydrodynamic equations follow from conservation of the stress
tensor, $\nabla_\mu T^{\mu\nu}=0$. This yields the relativistic
continuity, Navier-Stokes and Young-Laplace equations,
\begin{eqnarray}
&& \hspace{-1.5cm}u^{\mu}\nabla_{\mu}\rho + (\rho+ P) \vartheta=
\zeta\vartheta^2-q^\mu a_\mu -\nabla_\mu q^\mu
+2\eta\sigma^{\mu\nu}\nabla_\mu u_\nu\,,
\label{continuity:diss} \\
&& \hspace{-1.5cm} (\rho+ P) a^\nu = -P^{\mu\nu} \nabla_\mu  P
+\zeta\lrp{P^{\mu\nu}\nabla_\mu\vartheta +\vartheta u^\mu\nabla_\mu
u^\nu}
 +2\eta\lrp{\nabla_\mu\sigma^{\mu\nu}-u^\nu\sigma^{\mu\alpha}\nabla_\mu
u_\alpha}   \nonumber \\
&& \hspace{0.8cm} -\lrp{q^\mu\nabla_\mu u^\nu +\vartheta q^\nu
+u^\mu\nabla_\mu q^\nu -q^\mu a_\mu u^\nu}\,, \qquad {\rm with}\quad
P^{\mu\nu}\equiv g^{\mu\nu}+u^\mu u^\nu\,,
 \label{NavierS:diss}  \\
 &&  \hspace{-1.5cm}
 \left[ P - \zeta \vartheta+2\eta \lrp{\frac{1}{2}\vartheta
 +u^\mu n^\alpha\nabla_\alpha n_\mu}\right]^<_>=\sigma K\,, \qquad {\rm
with}\quad  K\equiv h_{\mu}^{\:\:\nu}\nabla_{\nu}  n^{\mu}\,.
 \label{YoungLap:diss}            % $K=h_{\mu}^{\:\:\nu}\nabla_{\nu} n^{\mu}$
 \end{eqnarray}
Here, $u^{\mu}$ is the fluid velocity, $\rho$, $P$, $\sigma$,
$\zeta$, $\eta$ and $\kappa$ are, respectively, the density,
pressure, surface tension, bulk viscosity, shear viscosity, and
thermal conductivity of the fluid. The quantities, $a^\mu$,
$\vartheta$, $\sigma^{\mu\nu}$ and $q^\mu$ are respectively the
acceleration, expansion, shear tensor, and heat flux.
$P^{\mu\nu}$ is the projector onto the hypersurface orthogonal to
$u^{\mu}$. The fluid boundary is defined by $f(x^{\mu})=0$, it has
unit spacelike normal $n_{\mu}=\partial_{\mu} f/|\partial f|$, and
$h^{\mu\nu}=g^{\mu\nu}-n^\mu n^\nu$ is the projector onto the
boundary. $K$ is the trace of its extrinsic curvature, and
$[Q]^<_>\equiv Q_<-Q_>$ is the jump on a quantity $Q$ when we cross
the boundary from the interior into the exterior of the plasma. In
the derivation of (\ref{YoungLap:diss}), the constraint that the
fluid velocity must be orthogonal to the boundary normal is used
(this guarantees that the fluid is confined inside the boundary),
\begin{eqnarray}
 u^\mu n_{\mu} = 0 \,. \label{constraintYL}
\end{eqnarray}

A fluid with local entropy density $s$ and local temperature ${\cal
T}$ satisfies the Euler relation and the Gibbs-Duhem relation given,
respectively, by
\begin{equation}
\rho+P ={\cal T} s\,, \qquad dP=sd{\cal T} \,, \label{Euler}
\end{equation}

In the SS dual system, we are interested in the long wavelength
limit of a Scherk-Schwarz compactification of a 4-dimensional CFT.
The $3d$ (non-conformal) plasma that results from the dimensional
reduction of the $4d$ conformal plasma has equation of state~\cite{Lahiri:2007ae}
\begin{equation}
 P=\frac{\rho-4\rho_0}{3}\; \Longleftrightarrow \; \rho+P=\frac{4}{3}\lp \rho-\rho_0\rp
 \,,\qquad
 s=4\alpha^{\frac{1}{4}}
\left(\frac{\rho-\rho_0}{3}\right)^{\frac{3}{4}}\,,\qquad {\cal
T}=\left(\frac{\rho-\rho_0}{3\,
\alpha}\right)^{\frac{1}{4}}\,.\label{ConfEqState}
 \end{equation}
with $\rho_0$ and $\alpha$  constants. This equation of state is
valid in or out of equilibrium and is normalized such that the
vacuum pressure vanishes.\footnote{The simplest possible plasma
configuration - a domain wall separating the fluid from the
vacuum - is in equilibrium when the plasma pressure vanishes.
This occurs for a critical temperature ${\cal T}_c=(\rho_0/\alpha)^{1/4}$~\cite{Lahiri:2007ae},
or equivalently for a critical fluid density $\rho_c=4\rho_0$ for which
$P=0$.}

%%%%%%%%%%%%%%%%%%%%%%%%%%%%%%%%%%%%%%%%%%%%%%%%%%%%%%%%%%%%%%%%%%%%%%
\subsection{Stationary plasmas \label{sec:stationary}}
%%%%%%%%%%%%%%%%%%%%%%%%%%%%%%%%%%%%%%%%%%%%%%%%%%%%%%%%%%%%%%%%%%%%%%

In this subsection we briefly review some general results derived in
\cite{Caldarelli:2008mv} which are valid for equilibrium plasmas. For
later sections, it is important to emphasize that these results are
valid not only for axisymmetric but also for non-axisymmetric plasma
configurations.

Stationary plasma configurations are both in hydrodynamical,
$u^\mu\nabla_\mu P=0$, and thermodynamical equilibrium
$u^\mu\nabla_\mu {\cal T}=0$. Equivalently, they have $\vartheta=0$,
$\sigma^{\mu\nu}=0$ and $q^\mu=0$. Consequently, the velocity field
of a stationary plasma must be a linear combination of the
background Killing vector fields ($\xi$ and $\chi$ are the
stationarity and rotational Killing vectors),
\begin{equation} u=\frac\T T \lp \xi+\Omega \chi\rp\,. \label{velocityfield}
\end{equation}
In particular this implies that a stationary plasma must be at
constant plasma temperature $T$ related to the local temperature
$\T$ by the Lorentz factor
\begin{equation} \gamma=\frac\T T= \left[-(\xi+\Omega\chi)^2\right]^{-1/2}\,,
\label{localtemp} \end{equation}
which is the redshift factor relating measurements done in the
laboratory and comoving frames. Combining the Euler relation
(\ref{Euler}) and the Young-Laplace equation (equation
(\ref{YoungLap:diss}) without the dissipative terms), we can relate
the plasma temperature $T$ to a combination of several magnitudes at
the fluid surface,
\begin{equation}\label{TsigmaK} T=\frac{\sigma K +\rho}{\gamma s}\,.
\end{equation}
In the plasma lump/black hole duality, $T$ maps to the Hawking temperature
of the horizon. A stationary plasma must also be rigidly rotating,
\ie with constant angular velocity $\Omega$, which in turn maps to the
horizon angular velocity in the duality.

Summarizing, the equations that a stationary fluid must satisfy are
given by \eqref{continuity:diss}-\eqref{YoungLap:diss} with
$\vartheta=0$, $\sigma^{\mu\nu}=0$ and $q^\mu=0$. In particular the
Young-Laplace equation \eqref{YoungLap:diss} reduces to\footnote{The
last term on the LHS of \eqref{YoungLap:Equil} vanishes in
equilibrium because use of $n^\mu n^\nu \sigma_{\mu\nu}=0$ implies
that $u^\mu n^\alpha\nabla_\alpha n_\mu=n^\mu n^\alpha\nabla_\alpha
u_\mu=0$.}
\begin{equation}
P_<-P_> =\sigma K\,.
 \label{YoungLap:Equil}
 \end{equation}
For the SS plasma in equilibrium, it further follows from
(\ref{localtemp}) and (\ref{ConfEqState}) that the pressure and
energy density satisfy the relations
\begin{equation}
 P=\frac{\rho_{*}}{3}\,\gamma^4-\rho_0\,,\qquad
\rho=\rho_{*}\,\gamma^4+\rho_0\,, \label{DpEquil}
\end{equation}
where $\rho_{*} \equiv 3\alpha T^4$ is a constant for any given plasma configuration.

%\begin{eqnarray}
%&& \hspace{-1.5cm}u^{\mu}\nabla_{\mu}\rho = 0\,,
%\label{continuity:Equil} \\
%&& \hspace{-1.5cm} (\rho+ P) a^\nu = -P^{\mu\nu} \nabla_\mu  P \,,
% \label{NavierS:Equil}  \\
%&& \hspace{-1.5cm}
% P_<-P_> =\sigma K\,,
% \label{YoungLap:Equil}
% \end{eqnarray}

%%%%%%%%%%%%%%%%%%%%%%%%%%%%%%%%%%%%%%%%%%%%%%%%%%%%%%%%%%%%%%%%%%%%%%%%%%%%%%%%%%%%%
\subsection{Conserved charges \label{sec:ConservedCharges}}
%%%%%%%%%%%%%%%%%%%%%%%%%%%%%%%%%%%%%%%%%%%%%%%%%%%%%%%%%%%%%%%%%%%%%%%%%%%%%%%%%%%%%

The constituent fluid of the plasma object has local energy density
$\rho$, pressure $P$, velocity $u^{\mu}$, local entropy density $s$,
and local temperature ${\cal T}$. These {\it local} quantities
provide the information we need to compute the thermodynamic
quantities (energy, angular momentum, entropy, temperature) of the
stationary configurations (plasma balls, plasma rings and plasma peanuts).

Here, we restrict our discussion to fluids that live in a
(2+1)-dimensional Minkowski background\footnote{We use polar
coordinates $(t,r,\phi)$, and the non-vanishing affine connections
are $\Gamma^r_{\phi\phi}=-r$ and
$\Gamma^\phi_{r\phi}=\Gamma^\phi_{\phi r}=1/r$.} with stationarity
timelike Killing vector $\xi=\partial_t$ and spacelike Killing
vector $\chi=\partial_\phi$ (for general results see
\cite{Caldarelli:2008mv}). We can then foliate the spacetime into
constant $t$ hypersufaces $\Sigma_t$ and $\xi^\mu$ is their unit
normal vector. Then, given any Killing vector $\psi^\mu$, one can
define the associated conserved charges ${\mathcal
Q}[\psi]=\int_{\Sigma_t}\!\!dV \;T_{\mu\nu}\xi^{\mu}\psi^\nu$, where
$dV$ is the induced volume measure on $\Sigma_t$. The fluid velocity
is given by (\ref{velocityfield}), \ie $ u^{\mu}=\gamma \lp
\delta^{\mu t} + \Omega\delta^{\mu\phi} \rp$ with  $\gamma=\lp
1-r^2\,\Omega^2 \rp^{-1/2}$. The energy and angular momentum of the
plasma associated, respectively, with the Killing vectors $\xi$ and
$\chi$ are then
\begin{eqnarray}
&& E[V]=\int_V\left[(\rho+P)(\xi\cdot u)^2+\lp \xi\cdot\xi\rp
P\right] dV-\sigma \int_{\Sigma_t} \xi^\mu\xi^\nu
h_{\mu\nu}|\p f|\delta(f)\,dV   \,,\nonumber\\
&& J[V]=\int_V\lp\rho+P\rp(\xi\cdot u)(\chi\cdot u)\,dV
-\sigma\int_{\Sigma_t} h_{\mu\nu}\xi^\mu\chi^\nu|\p f|\delta(f)dV
\,.
 \label{Var:charges}
\end{eqnarray}
Note that for axisymmetric solutions (plasma balls and plasma rings)
the boundary term in $J$ proportional to $\sigma$ vanishes. It is
however present for non-axisymmetric solutions where $\chi\cdot
n\neq 0$ (\ie when the fluid boundary is not invariant under the
action of $\chi$) \cite{Caldarelli:2008mv}.
The total entropy of the fluid is the conserved charge associated to
the entropy density current $su^\mu$,
\begin{eqnarray}
S=-\int_V dV\,(\xi\cdot u) s =\int_V\, dV \gamma s
\,.\label{Var:Entropy}
\end{eqnarray}

%%%%%%%%%%%%%%%%%%%%%%%%%%%%%%%%%%%%%%%%%%%%%%%%%%%%%%%%%%%%%%%%%%%%%%%%%%%%%%%%%%%%%%%%%%%%%%%%%%%%%%%%%%%%%%%%%%%%%%%%
\setcounter{equation}{0}
\section{Stationary solutions: balls, rings and lobed plasmas \label{sec:ballsRingsLobed}}
%%%%%%%%%%%%%%%%%%%%%%%%%%%%%%%%%%%%%%%%%%%%%%%%%%%%%%%%%%%%%%%%%%%%%%%%%%%%%%%%%%%%%%%%%%%%%%%%%%%%%%%%%%%%%%%%%%%%%%%%

In this section we are interested in rigidly rotating equilibrium
configurations in a $(2+1)$-dimensional Minkowski background. In
addition to the two families of axisymmetric solutions composed of
plasma balls and plasma rings there exist also non-axisymmetric
configurations or $m$-lobed plasmas, where $m$ runs over the set of
positive integers excluding 1. The plasma balls and rings were
discussed in detail in \cite{Lahiri:2007ae}. Because we will later
study the phase diagram of these solutions we review here their
properties, following closely \cite{Lahiri:2007ae}.

%%%%%%%%%%%%%%%%%%%%%%%%%%%%%%%%%%%%%%%%%%%%%%%%%%%%%%%%%%%%%%%%%%%%%%%%%%%%%%%%%%%%%
\subsection{Axisymmetric solutions: plasma balls and plasma rings \label{sec:ballsRings}}
%%%%%%%%%%%%%%%%%%%%%%%%%%%%%%%%%%%%%%%%%%%%%%%%%%%%%%%%%%%%%%%%%%%%%%%%%%%%%%%%%%%%%

Consider plasma configurations in a $3d$  Minkowski background
parametrized by coordinates $(t,r,\phi)$. The axisymmetry
requirement demands that the boundaries of the plasma depend only on
$r$. Each boundary is thus defined by the condition ($\alpha$
specifies a particular boundary in the case where more than one is
present)
\begin{equation} f(r)= r-R_\alpha=0\,,
\end{equation}
and has unit normal $n_\mu=\frac{\partial_\mu f}{|\partial
f|}=\delta_{\mu r}$.  Its extrinsic curvature is $K=\frac{1}{R_\alpha}$.
Following \cite{Lahiri:2007ae}, it is convenient to frame our
discussion in terms of the dimensionless variables,
\begin{equation}\label{newvars}
    \tO = \frac{\sigma\Omega}{\rho_0} \,,   \qquad
    \z = \frac{\rho_0 r}{\sigma} \,,   \qquad
    v = \Omega r = \tO \z \,,
\end{equation}
and also to use dimensionless thermodynamic quantities,
\begin{equation}\label{redThermo}
  \tE = \frac{\rho_0 E}{\pi\sigma^2} \,,\quad
  \tJ = \frac{\rho_0^2 J}{\pi\sigma^3} \,,\quad
  \tS = \frac{\rho_0^{5/4}S}{\pi\alpha^{1/4}\sigma^2} \,,\quad
  \tT = T\lrp{\frac{\alpha}{\rho_0}}^{1/4}\,.
\end{equation}

In $3d$ the possible axisymmetric stationary configurations are the
plasma balls and the plasma rings:

\begin{itemize}
\item
Plasma balls are characterized by having a single axisymmetric outer
surface at $r=\ro$ and by $P_>=0$. Its properties can be found using
the equation of state (\ref{DpEquil}), and the Young-Laplace
equation (\ref{YoungLap:Equil}). It follows that plasma balls must
satisfy the condition (with $0\leq \vo \leq 1$)
\begin{equation}\label{TildeRHO:rho0} \frac{\rho_{*}}{3\rho_0}= \lrp{1+\frac{\tO}{\vo}}\lrp{1-\vo^2}^{2} \,.
\end{equation}

\item Plasma rings have an axisymmetric inner surface at $r=\ri$ (where
$P_<=0$), in addition to the outer surface at $r=\ro$ (where
$P_>=0$). The equation of state (\ref{DpEquil}), and the
Young-Laplace equation (\ref{YoungLap:Equil}) introduce a constraint
equation for these objects,
\begin{equation}\label{Ring:gConstraint}
 \lrp{1+\frac{\tO}{\vo}}\lrp{1-\vo^2}^{2} = \lrp{1-\frac{\tO}{\vi}}\lrp{1-\vi^2}^{2} \,.
\end{equation}
It constrains the three variables $\vo$, $\vi$ and $\tO$ as, \eg
$\vi=\vi(\vo,\tO)$. An inspection of it concludes that there is a
minimum $\vo$, call it $\vo^*$, above which \eqref{Ring:gConstraint}
is valid \cite{Lahiri:2007ae}. So, plasma rings exist only for $\vo
\geq \vo^*$. In fact there are two families of black rings. One is
called the fat plasma ring and exists for $\tO \leq\vi \leq\vi^*$
(where $\vi^*<\vo^*$ is such that the derivative of of the RHS of
\eqref{Ring:gConstraint} w.r.t. $\vi$ vanishes). The second, dubbed
as thin plasma ring, exists for $\vi^* \leq\vi \leq 1$. At
$\vi=\vi^*$ the two families meet at a regular solution. The
condition $\vi \geq \tO$ guarantees that the temperature is
non-negative.
\end{itemize}

The dimensionless energy, angular momentum and entropy of the plasma
rings are, respectively,
\begin{eqnarray}\label{Ring:charges}
    \tE &=& \frac{4(\vo^2-\vi^2) - (\vo^4-\vi^4)
                 + 5\tO (\vo+\vi) - \tO (\vo^3+\vi^3)}
             {\tO^2}\,, \nonumber\\
  \tJ &=& \frac{2(\vo^4-\vi^4)+2\tO (\vo^3+\vi^3)}{\tO^3} \,, \nonumber\\
  \tS &=& \frac{4}{\tO^2}
        \lrpp{\vo^2\sqrt{1-\vo^2}\lrp{1+\frac{\tO}{\vo}}^{3/4}
            -\vi^2\sqrt{1-\vi^2}\lrp{1-\frac{\tO}{\vi}}^{3/4}}\,,
\end{eqnarray}
and the corresponding charges for the plasma balls can be obtained
from these formulas by taking $\vi=0$.

%%%%%%%%%%%%%%%%%%%%%%%%%%%%%%%%%%%%%%%%%%%%%%%%%%%%%%%%%%%%%%%%%%%%%%%%%%%%%%%%%%%%%
\subsection{Plasma peanuts and $m$-lobed configurations
\label{sec:peanuts}}
%%%%%%%%%%%%%%%%%%%%%%%%%%%%%%%%%%%%%%%%%%%%%%%%%%%%%%%%%%%%%%%%%%%%%%%%%%%%%%%%%%%%%

%%%%%%%%%%%%%%%%%%%%%%%%%%%%%%%%%%%%%%%%%%%%%%%%%%%%%%%%%%%%%%%%%%%%%%%%%%%%%%%%%%%%%
\subsubsection{Boundary equation for lobed plasmas \label{sec:peanutsEqs}}
%%%%%%%%%%%%%%%%%%%%%%%%%%%%%%%%%%%%%%%%%%%%%%%%%%%%%%%%%%%%%%%%%%%%%%%%%%%%%%%%%%%%%

According to the discussion in Section~\ref{sec:stationary} the
stationarity restriction can accommodate non-axisymmetric
configurations as long as they are in rigid rotation.  Taking this
into account, we drop the assumption of axisymmetry and define the
surface of the plasma lump through the condition
\be f(t,r,\phi) \equiv r - R(t-\phi/\Omega) = 0\,.
\label{ansatz:RigRot} \ee
The boundary's unit normal is
 \be n_{\mu}=|\partial f|^{-1}\lp -R'\delta_{\mu}^{\:t}+
\delta_{\mu}^{\:r}+\frac{R'}{\Omega}\delta_{\mu}^{\:\phi} \rp
\,,\qquad  |\partial f|=\lp \frac{\Omega^2 R^2 +R^{\prime\,2}\lp
1-\Omega^2 R^2\rp}{\Omega^2 R^2}\rp^{\frac{1}{2}},
 \label{normal}
 \ee
where $R'=\frac{dR(x)}{dx}$ with $x=t-\phi/\Omega$.  The condition
that no fluid flows through the surface, namely $u^\mu n_\mu = 0$,
is manifestly satisfied by the ansatz~(\ref{ansatz:RigRot}) when the
velocity field takes the form $u^\mu=\gamma(\delta^\mu_t
+\Omega\delta^\mu_\phi)$.  The trace of the extrinsic curvature of
the boundary in the general rigid rotation case is
\be K = \Omega \frac{-R R''\lp 1-\Omega^2 R^2\rp +
 R'^2 \lp 2-\Omega^2 R^2\rp + \Omega^2 R^2}
{\lpp \Omega^2 R^2 +R'^2 \lp 1-\Omega^2 R^2\rp \rpp^{3/2}} \ee

Introducing the dimensionless quantities~\eqref{newvars} and
\begin{equation}\label{newvarsII}
    \psi= -\Omega x=\phi-\Omega t \,, \qquad
    \vo(\psi)=\Omega R(x)\,,   \qquad
    \vo'(\psi)=R'(x)\,, \qquad k=
    \frac{\rho_*}{3\rho_0}\,.
\end{equation}
the Young-Laplace equation~\eqref{YoungLap:diss}, after
employing~\eqref{ConfEqState} and~\eqref{localtemp}, becomes
\begin{equation}\label{YL:profiles}
  \frac{\vo \vo'' (1-\vo^2)-\vo'^{\,2}(2-\vo^2)-\vo^2}{\lpp
\vo^2+\vo'^{\,2}(1-\vo^2) \rpp^{3/2}}
  +\frac{1}{\tO}\lpp k(1-\vo^2)^{-2}-1\rpp =0\,.
\end{equation}
This equation admits a first integral, namely the quantity
\be \label{1stIntegral} \frac{\vo^2}{
\sqrt{\vo^2+\vo'^{\,2}(1-\vo^2)} }
  -\frac{1}{2\tO}\lpp k(1-\vo^2)^{-1}-\vo^2\rpp \equiv Q
\ee
is a constant independent of $\psi$.

Now consider the stress tensor for a perfect fluid with surface
tension $\sigma$.  The relevant $T_{\mu\nu}$ components to compute
the charges are
\begin{eqnarray}
&&T_{tt}=\gamma^2 \lp \rho+\Omega^2 r^2 P\rp
\Theta(-f)+\frac{\sigma}{\Omega R} \frac{\Omega^2 R^2+
R^{\prime\,2}}{\sqrt{\Omega^2 R^2+ R^{\prime\,2}\lp 1-\Omega^2
R^2\rp}}\,\delta(f)
\,,\nonumber\\
&& T_{t\phi}=-\gamma^2 \Omega r^2 \lp \rho+P\rp \Theta(-f)-\sigma
\frac{R R^{\prime\,2}}{\sqrt{\Omega^2 R^2+ R^{\prime\,2}\lp
1-\Omega^2 R^2\rp}} \,\delta(f) \,.\label{GenLumpTuv}
\end{eqnarray}
The induced measure on a constant $t$ hypersurface $\Sigma_t$ used
to compute the conserved charges in \eqref{Var:charges} and
\eqref{Var:Entropy} is $dV=r dr d\phi
=\frac{\sigma^2}{\rho_0^2\tO^2}\,v dv d\psi$. Using
\eqref{Var:charges} and \eqref{Var:Entropy}, the energy, angular
momentum and entropy of the lobed plasma are then
\begin{eqnarray}
   \tE &=&\frac{1}{2\pi\tO^2}\int_0^{2\pi} d\psi \lpp
\vo^2 + k\,\frac{\vo^2(3-\vo^2)}{\lp 1-\vo^2 \rp^2}
+2\tO\frac{\vo^2+\vo'^2}
   {\sqrt{\vo^2+\vo'^2(1-\vo^2)}}\rpp \,, \nonumber\\
   \tJ &=& \frac{1}{\pi\tO^3} \int_0^{2\pi} d\psi \lpp  \frac{k\vo^4}{\lp
1-\vo^2 \rp^2}
   + \frac{\tO \vo^2\vo'^2}{\sqrt{\vo^2+\vo'^2(1-\vo^2)}} \rpp\,, \nonumber\\
  \tS &=& \frac{2}{\pi \tO^2}\int_0^{2\pi} d\psi \, \frac{k^{3/4}\vo^2}{
1-\vo^2}\,, \label{ChargesPeanut}
\end{eqnarray}
where $\vo=\vo(\psi)$ is the velocity of the plasma peanut boundary
that solves the Young-Laplace equation and
$\vo'=\frac{d\vo}{d\psi}$. Note that by setting
$\vo'(\psi)=0$, and $k=\lrp{1+\tO/\vo}\lrp{1-\vo^2}^{2}$ (see
\eqref{TildeRHO:rho0}) we get the plasma ball charges (equation
\eqref{Ring:charges} with $\vi=0$).

Lobed plasma configuration are those whose boundary profiles satisfy
equation \eqref{YL:profiles} (or \eqref{1stIntegral}). Next we find
these solutions in a perturbative expansion around the plasma ball
bifurcation point.

%%%%%%%%%%%%%%%%%%%%%%%%%%%%%%%%%%%%%%%%%%%%%%%%%%%%%%%%%%%%%%%%%%%%%%%%%%%%%%%%%%%%%
\subsubsection{Lobed plasmas from perturbed analysis of plasma balls \label{sec:peanutsPert}}
%%%%%%%%%%%%%%%%%%%%%%%%%%%%%%%%%%%%%%%%%%%%%%%%%%%%%%%%%%%%%%%%%%%%%%%%%%%%%%%%%%%%%

As a warm-up exercise that will also be very useful to check our
numerical results against, we find lobed plasma configurations by
performing a perturbative analysis of the plasma balls around the
bifurcation point to the new phase.

Consider a small non-axisymmetric perturbation of the plasma ball,
\begin{eqnarray} && \vo(\psi) =\widehat{v}_\mathrm{o} \lpp 1+\varepsilon\, \nu(\psi) +
\mathcal{O}(\varepsilon^2)\rpp \,, \nonumber\\
&& \tO=\widehat{\Omega}+\varepsilon \omega+
\mathcal{O}(\varepsilon^2)\,, \qquad  k=\widehat{k}+\varepsilon
\kappa + \mathcal{O}(\varepsilon^2)\,. \label{Ball:Perturb}
\end{eqnarray}
where the hat refers to the (constant) unperturbed quantities,
$\varepsilon\ll 1$ is the expansion parameter, and the unperturbed
plasma ball radius $\widehat{v}_\mathrm{o}$ necessarily obeys
\eqref{TildeRHO:rho0}, which we write here as
\be \label{constraint1} \widehat{\Omega} = \widehat{v}_\mathrm{o}
\lpp \widehat{k}(1-\widehat{v}_\mathrm{o}^2)^{-2} -1 \rpp \,. \ee
Note that a consistent perturbation requires that we disturb not
only $\vo$ but also $\tO$ and $k$, as done in \eqref{Ball:Perturb}.
Linearizing the Young-Laplace equation~\eqref{YL:profiles} we then
get
\be \label{perturb:eigen} \nu'' + m^2 \nu = \Delta \,,  \ee
with
\begin{equation} \label{Ball:m}
m^2 = \frac{1}{1-\widehat{v}_\mathrm{o}^2} + \frac{4\,\widehat{k} \,\widehat{v}_\mathrm{o}^3}{\widehat{\Omega}
\lp 1-\widehat{v}_\mathrm{o}^2\rp^4} \,, \qquad \Delta=
\widehat{v}_\mathrm{o}\,
 \frac{\lpp \widehat{k}-\lp 1-\widehat{v}_\mathrm{o}^2\rp^2 \rpp \omega-\widehat{\Omega} \, \kappa}{\widehat{\Omega}^2\lp 1-\widehat{v}_\mathrm{o}^2\rp^3 }\,.
\end{equation}

The solution for the perturbation is oscillatory,
\be \nu(\psi) = A \sin(m\psi+\beta) +\Delta/m^2\,, \label{OscSolution} \ee
and imposing periodicity $\psi \rightarrow \psi + 2\pi$ implies that
$m\in\bZ$.  The plasma peanuts correspond to $m=2$.
The definition of the small expansion parameter,
\begin{equation}
\varepsilon \equiv \frac{1}{\widehat{v}_\mathrm{o}} \int^\pi_0 d\psi \lpp \vo(\psi)-\widehat{v}_\mathrm{o} \rpp \nu(\psi) \,,
\end{equation}
provides a normalization for the function $\nu(\psi)$:
\begin{equation}
\int^\pi_0 d\psi \lpp\nu(\psi)\rpp^2 = 1 \,.
\end{equation}
This in turn may be used to fix the amplitude $A$ as
\begin{equation}
A = \sqrt{\frac{2}{\pi} \left( 1-\frac{\pi \Delta^2}{m^4} \right) } \,.
\end{equation}
However, $A$ does not enter the first order expansions of the
various thermodynamic quantities so we shall not need it.

Note that combining eqs.~\eqref{constraint1} and \eqref{Ball:m} we
obtain the constraint
\begin{equation}
m^2 = \frac{1 + 3\widehat{v}_\mathrm{o}^2 +
4\widehat{v}_\mathrm{o}^3/\widehat{\Omega}}{(1-\widehat{v}_\mathrm{o}^2)^2}
\,. \label{constraint2}
\end{equation}
Recall that for a given energy, the unperturbed plasma ball
quantities $\widehat{v}_\mathrm{o}$ and $\widehat{\Omega}$ are
related by the first equation in \eqref{Ring:charges}. Relation
\eqref{constraint2} then determines the value of
$\widehat{v}_\mathrm{o}$ (or $\widehat{\Omega}$) where the $m$-lobed
plasma bifurcates from the plasma ball branch for a given energy. In
the non-relativistic limit, $\widehat{v}_\mathrm{o}\ll 1$ and
$\frac{\sigma}{\rho_0 R_\mathrm{o}}\ll 1$, eq. (\ref{constraint2})
reduces to $4R_\mathrm{o}^3\Omega^2\rho_0/\sigma=m^2-1$ and agrees
with the bifurcation point found in
\cite{Cardoso:2009bv}.\footnote{There is an interesting feature
associated with this value. In \cite{Cardoso:2009bv} the classical
critical rotation was found studying perturbations of the
dissipative hydrodynamic
equations~\eqref{continuity:diss}-\eqref{YoungLap:diss} in their
non-relativistic limit. Curiously, if we repeat the perturbation
analysis starting this time with the {\it non}-dissipative fluid
equations we find that the value of the critical rotation {\it
shifts} to $4R_\mathrm{o}^3\Omega^2\rho_0/\sigma=m(m+1)$. So, even a
vanishingly small viscosity has a dramatic effect on the critical
rotation at which instability sets in. Viscosity introduces a
singular-limit, where the ``limit of the theory is not the theory of
the limit''. In classical fluids this property is very familiar and
well studied \cite{ViscClassic,Scriven1991}.}

Having identified the bifurcation point we can now determine the
slope of the new phase branch curve in the neighborhood of the
bifurcation point.  Expanding to first order the dimensionless
energy, angular momentum and entropy of the lobed plasmas (given by
eq.~(\ref{ChargesPeanut})) around the plasma balls, we get
\begin{eqnarray}
\tE &=& \tE_{pb} + \varepsilon \frac{\widehat{v}_\mathrm{o}^2}{\widehat{\Omega}^2}
        \left\{ \frac{3-\widehat{v}_\mathrm{o}^2}{(1-\widehat{v}_\mathrm{o}^2)^2}\kappa -
        \frac{2}{\widehat{\Omega}} \left[ \lp 1+\frac{\widehat{\Omega}}{\widehat{v}_\mathrm{o}} \rp +
        \frac{\widehat{k}(3-\widehat{v}_\mathrm{o}^2)}{(1-\widehat{v}_\mathrm{o}^2)^2} \right] \omega +
        \left[ \lp 1+\frac{\widehat{\Omega}}{\widehat{v}_\mathrm{o}} \rp +
        \frac{\widehat{k}(3+\widehat{v}_\mathrm{o}^2)}{(1-\widehat{v}_\mathrm{o}^2)^3} \right]
        \frac{2\Delta}{m^2} \right\} \nonumber\\
    &&  + O(\varepsilon^2) \,, \nonumber\\
\tJ &=& \tJ_{pb} + \varepsilon \, \frac{\widehat{k}\, \widehat{v}_\mathrm{o}^4}{\widehat{\Omega}^3 (1-\widehat{v}_\mathrm{o}^2)^2}
        \left\{\frac{2\kappa}{\widehat{k}} - \frac{6\omega}{\widehat{\Omega}} + \frac{8\Delta}{m^2(1-\widehat{v}_\mathrm{o}^2)} \right\}
        + O(\varepsilon^2) \,, \nonumber\\
\tS &=& \tS_{pb} + \varepsilon \, \frac{\widehat{k}^{3/4} \widehat{v}_\mathrm{o}^2}{\widehat{\Omega}^2 (1-\widehat{v}_\mathrm{o}^2)}
        \left\{\frac{3\kappa}{\widehat{k}} - \frac{8\omega}{\widehat{\Omega}} + \frac{8\Delta}{m^2(1-\widehat{v}_\mathrm{o}^2)} \right\}
        + O(\varepsilon^2) \,.
\label{ChargesPeanutPert}
\end{eqnarray}
Here, the subscript {\it pb} is appended to quantities pertaining to
plasma balls (see eq.~\eqref{Ring:charges} for their definitions).

In the next subsection we shall obtain the phase diagram at fixed
energy. Since we focus on 2-lobed perturbations of plasma balls, we
require that the energy $\tE$ of the non-axisymmetric configuration
actually equals $\tE_{pb}$.  Thus, the energy constraint reads
\begin{equation}
\frac{3-\widehat{v}_\mathrm{o}^2}{(1-\widehat{v}_\mathrm{o}^2)^2}\kappa -
\frac{2}{\widehat{\Omega}} \left[ \lp 1+\frac{\widehat{\Omega}}{\widehat{v}_\mathrm{o}} \rp +
\frac{\widehat{k}(3-\widehat{v}_\mathrm{o}^2)}{(1-\widehat{v}_\mathrm{o}^2)^2} \right] \omega +
\left[ \lp 1+\frac{\widehat{\Omega}}{\widehat{v}_\mathrm{o}} \rp +
\frac{\widehat{k}(3+\widehat{v}_\mathrm{o}^2)}{(1-\widehat{v}_\mathrm{o}^2)^3} \right]
\frac{2\Delta}{m^2} = 0 \,.
\label{constraint:Energy}
\end{equation}
Together with \eqref{Ball:m}, the energy constraint yields a linear relation
between the first-order perturbation quantities,
\begin{equation}
\kappa = \omega\, \frac{(1-\widehat{v}_\mathrm{o}^2)^2
(9-39\widehat{v}_\mathrm{o}^2+54\widehat{v}_\mathrm{o}^4-27\widehat{v}_\mathrm{o}^6+4\widehat{v}_\mathrm{o}^8)}
{\widehat{v}_\mathrm{o}(-3+10\widehat{v}_\mathrm{o}^2-8\widehat{v}_\mathrm{o}^4+2\widehat{v}_\mathrm{o}^6)} \,,
\end{equation}
where we have used eqs.~\eqref{constraint1} and \eqref{constraint2} to
eliminate $\widehat{k}$ and $\widehat{\Omega}$ in favor of
$\widehat{v}_\mathrm{o}$ (and replaced $m=2$).

At this point we have not yet fully determined the first-order
quantities $\{ \nu(\psi), \omega, \kappa \}$, but to compute the
slope at which the plasma peanut branches off from the plasma ball
curve in the $\tS-\tJ$ phase diagram we need not go
further\footnote{In the non-relativistic case the perturbative
analysis must be extended to second order~\cite{Cardoso:2009bv}.}.
The reason for this is that the slope (at the bifurcation point) is
given by
\begin{equation}
\lp \frac{\p \tS}{\p \tJ} \rp^\mathrm{peanut}_{\tE}
= \lim_{\varepsilon\rightarrow 0} \frac{\lp \frac{\p \tS}{\p \varepsilon} \rp_{\tE}}{\lp \frac{\p \tJ}{\p \varepsilon} \rp_{\tE}} \,,
\end{equation}
and therefore the only terms for $\tS$ and $\tJ$ in
eq.~\eqref{ChargesPeanutPert} that contribute are those linear in
$\varepsilon$. Taking the ratio, the undetermined first-order
quantity that was left cancels out, leaving an expression as a
function of $\widehat{v}_\mathrm{o}$ only.  The final result for the
slope of the plasma peanut at the bifurcation point in the
$\tS(\tJ)$ phase diagram is
\begin{equation}
\lp \frac{\p \tS}{\p \tJ} \rp^\mathrm{peanut}_{\tE}
= - \frac{4\widehat{v}_\mathrm{o}^3}{\lp 3-4\widehat{v}_\mathrm{o}^2 \rp^{1/4}
\left[ (1-\widehat{v}_\mathrm{o}^2)(3-11\widehat{v}_\mathrm{o}^2+4\widehat{v}_\mathrm{o}^4) \right]^{3/4}} \,.
\label{slope:SvsJ}
\end{equation}
For $\tE=40$, eqs.~(\ref{Ring:charges}) and~(\ref{constraint2})
determine the bifurcation point to occur at
$\widehat{v}_\mathrm{o} \simeq 0.378$ ($\widehat{\Omega} \simeq 0.14$).
Plugging this value into eq.~\eqref{slope:SvsJ} we obtain
$\p\tS/\p\tJ \simeq -0.142$. This perturbative result will be
compared with the numerical result of the next subsection and a good
agreement will be found.

%%%%%%%%%%%%%%%%%%%%%%%%%%%%%%%%%%%%%%%%%%%%%%%%%%%%%%%%%%%%%%%%%%%%%%%%%%%%%%%%%%%%%
\subsubsection{Phase diagram of stationary plasmas \label{sec:numerics}}
%%%%%%%%%%%%%%%%%%%%%%%%%%%%%%%%%%%%%%%%%%%%%%%%%%%%%%%%%%%%%%%%%%%%%%%%%%%%%%%%%%%%%

\begin{figure}[ht]
\begin{tabular}{lcl}
\epsfig{file=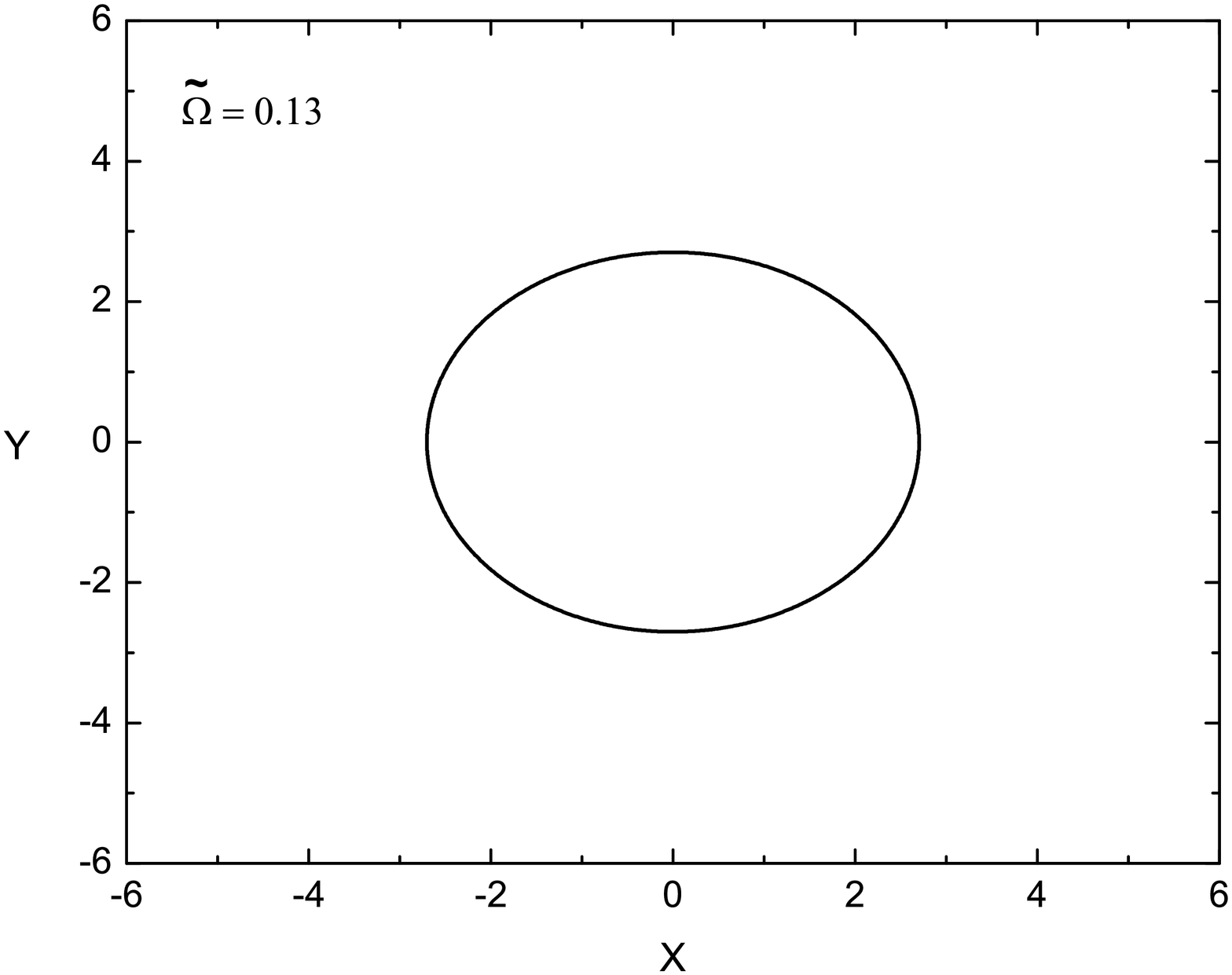,width=5cm,angle=0}&\epsfig{file=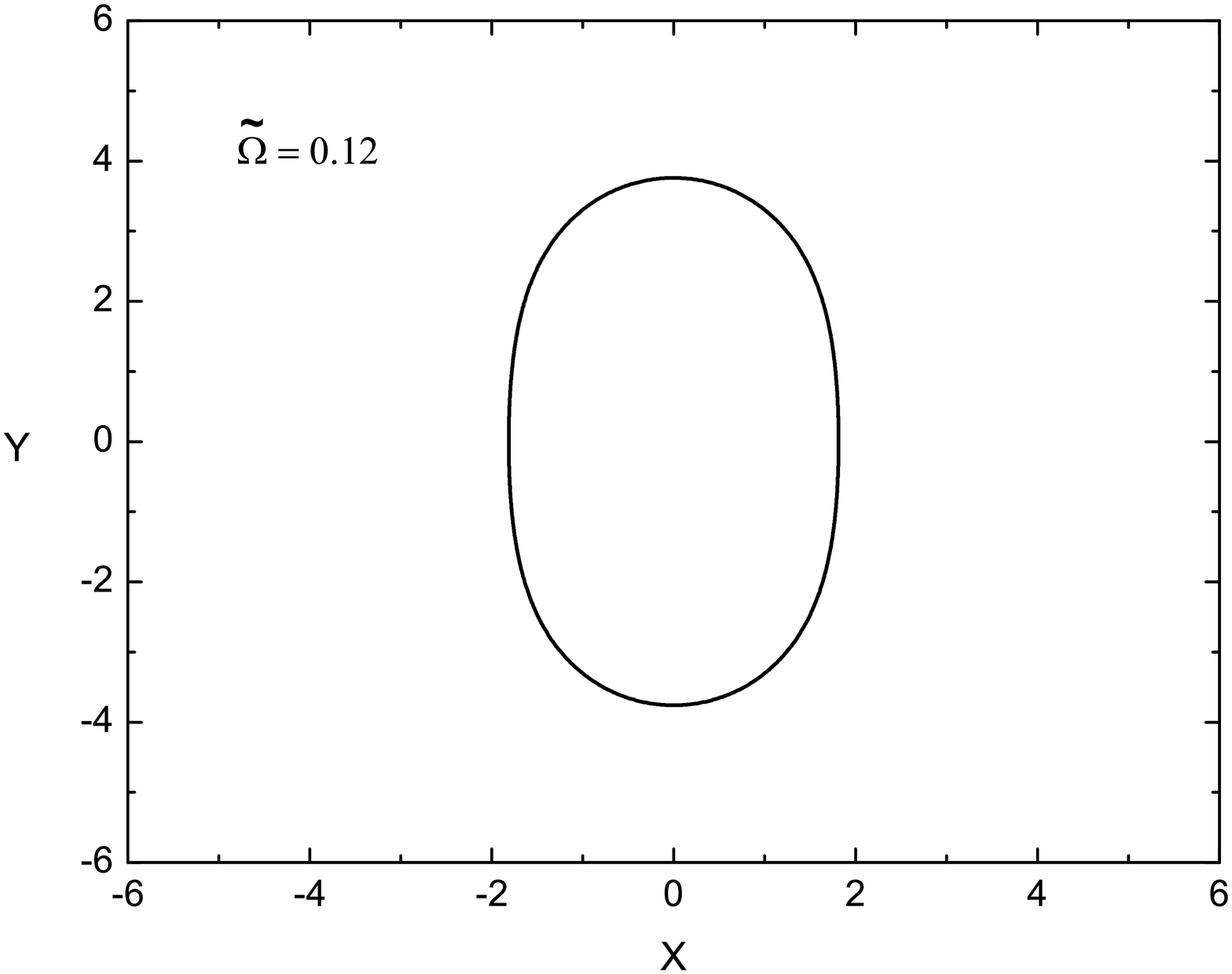,width=5cm,angle=0}&\epsfig{file=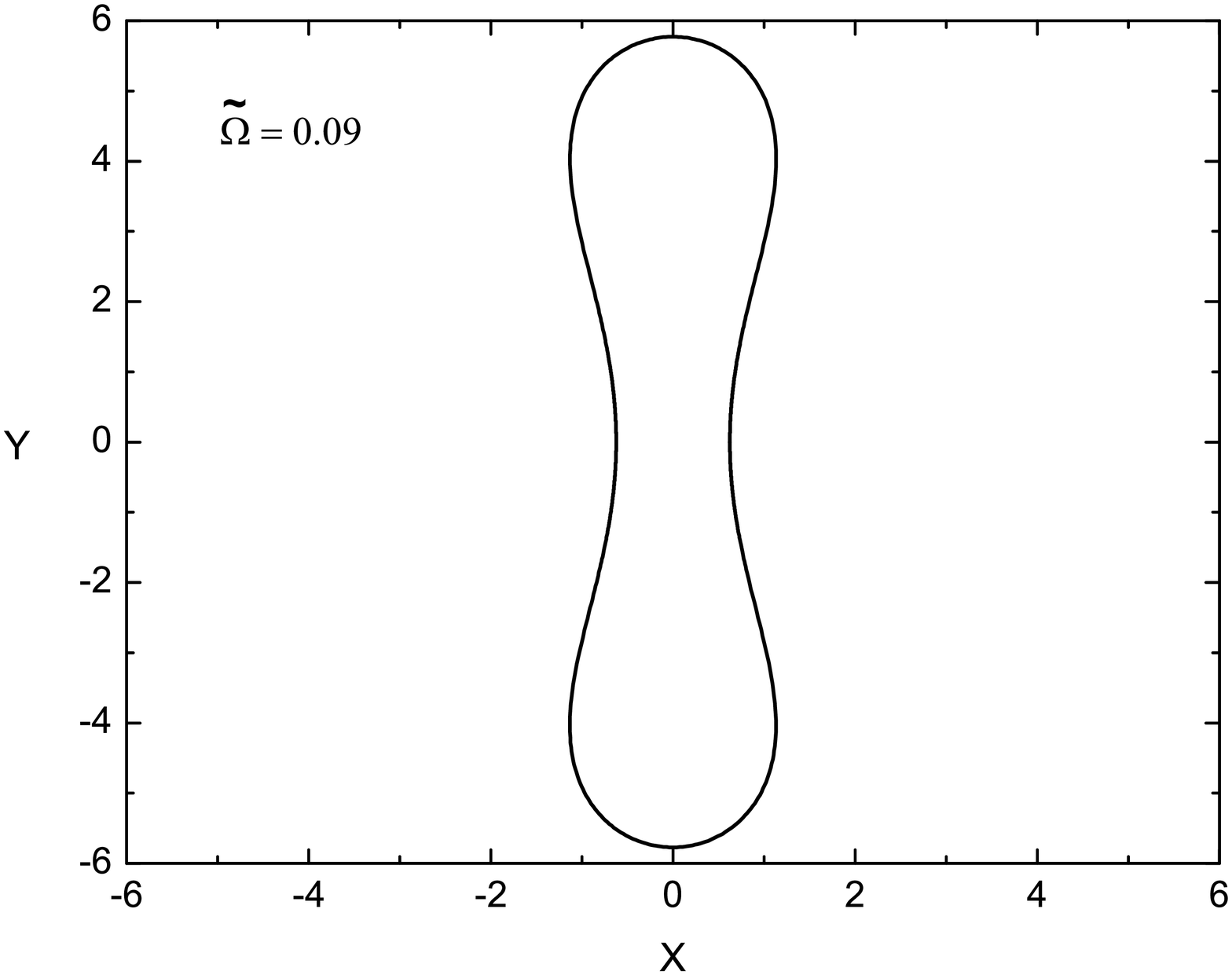,width=5cm,angle=0}
\end{tabular}
\caption{Different configurations of a two-lobed plasma for three
different values of angular velocity $\tO=0.13\,,0.12\,,0.09$. The
energy is kept fixed at $\tilde{E}=40$ and the bifurcation rotation
occurs at $\tO\simeq 0.16$. (The rotation axis is normal to the
page). \label{fig:shape}}
\end{figure}

\begin{figure}[ht]
\begin{tabular}{ccc}
\epsfig{file=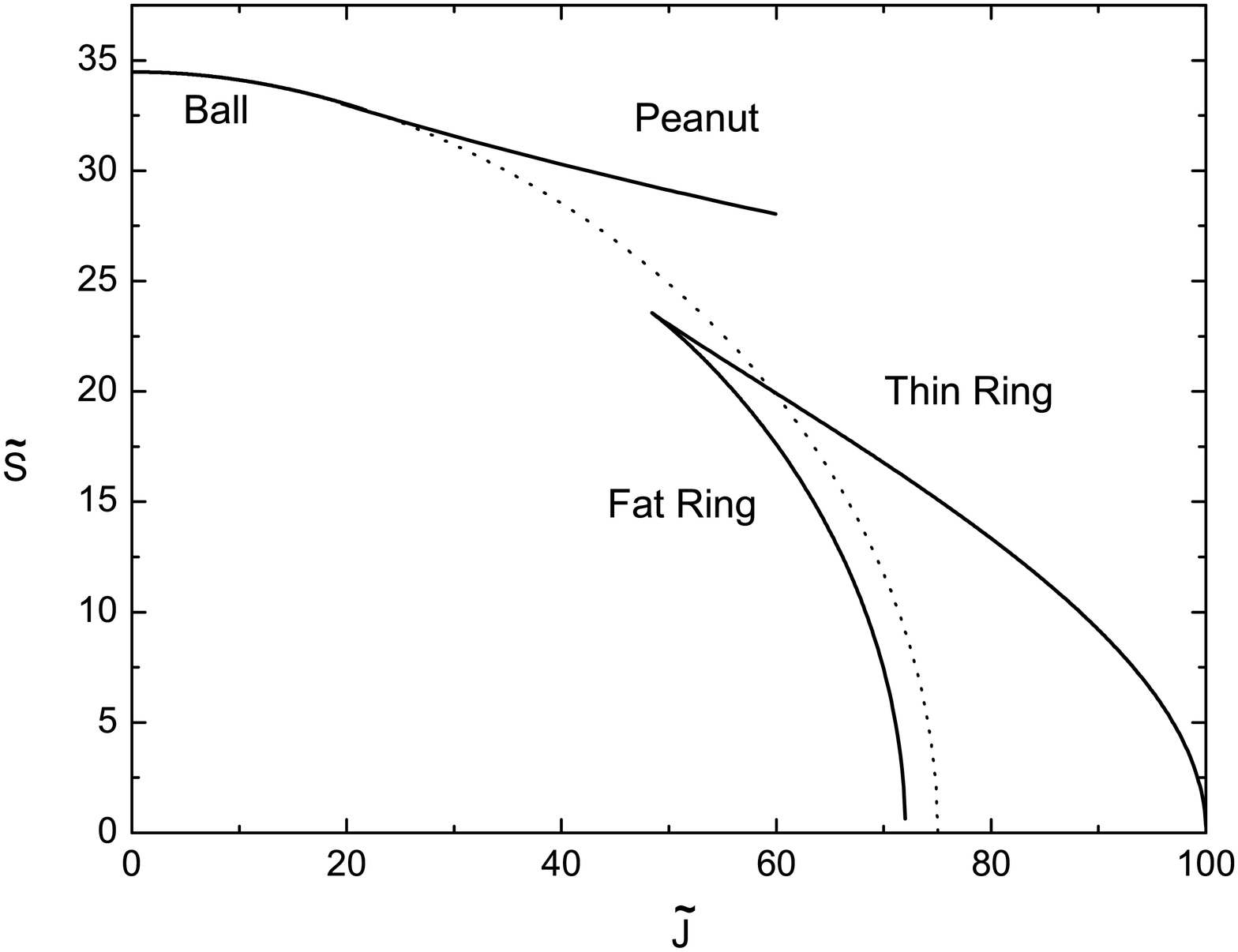,width=5cm,angle=0}&\epsfig{file=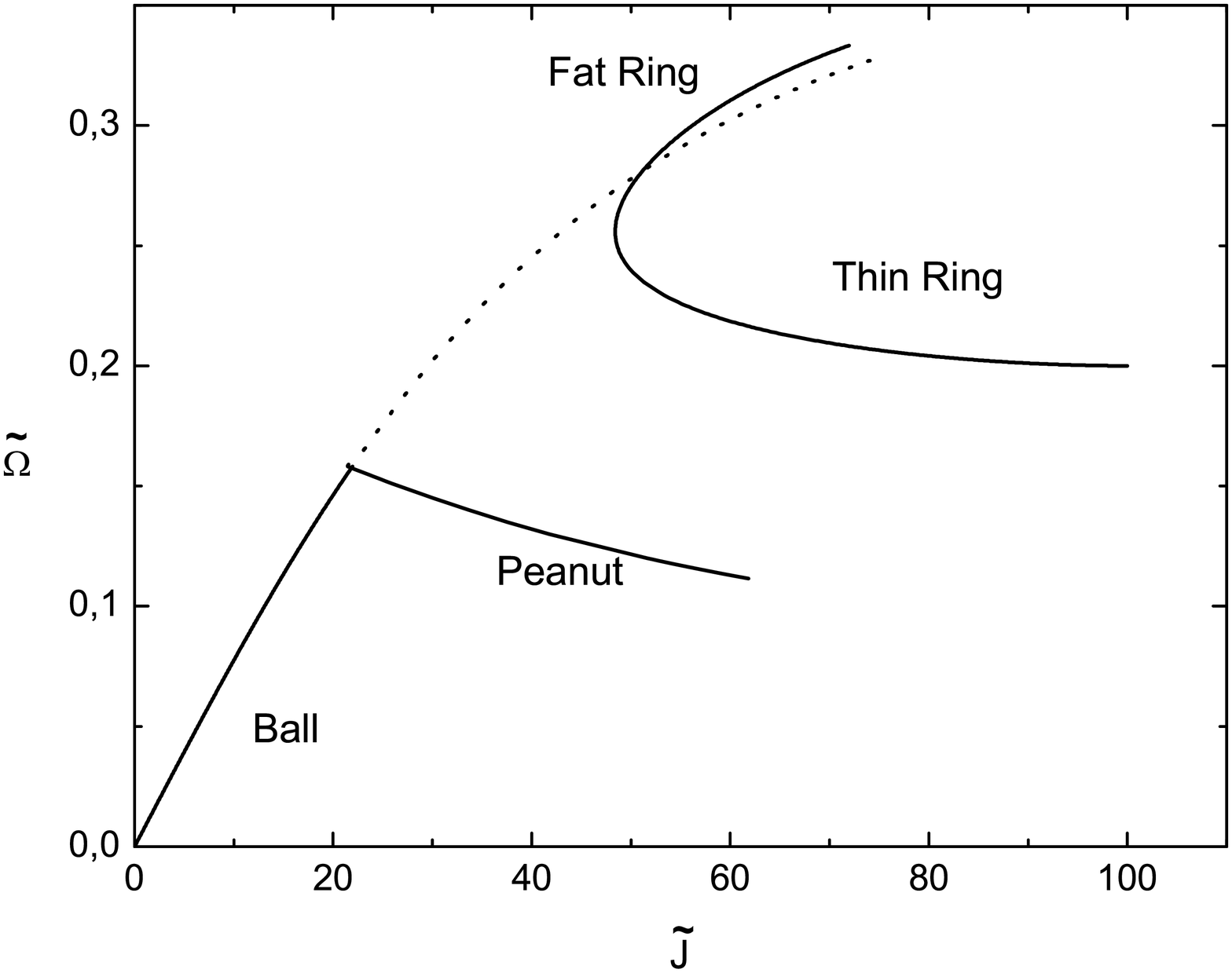,width=5cm,angle=0}&\epsfig{file=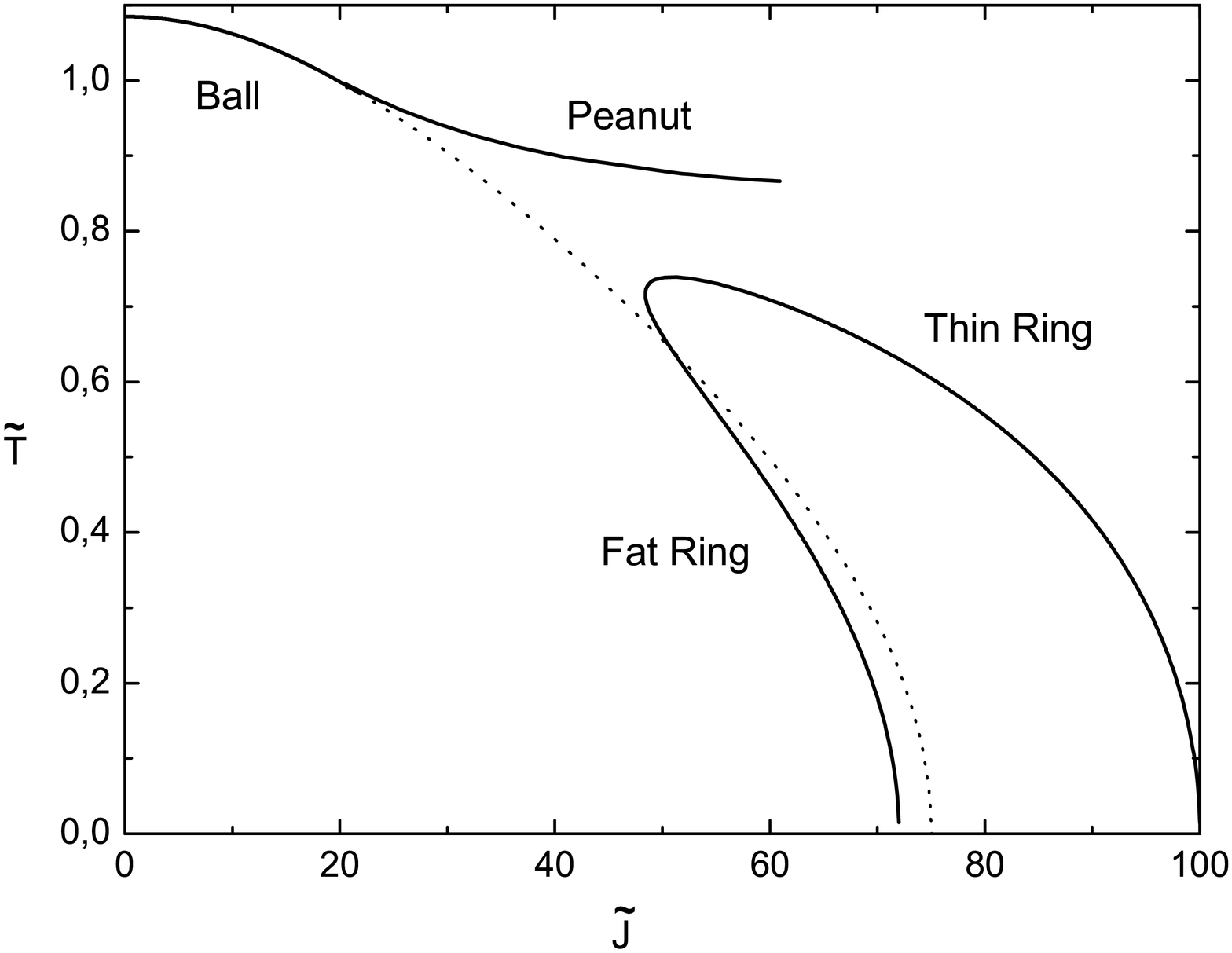,width=5cm,angle=0}
\end{tabular}
\caption{Phase diagrams $\tS(\tJ)$, $\tO(\tJ)$ and $\tilde{T}(\tJ)$
with the two-lobed, ``peanut-like'' configurations, along with the
plasma balls, and thin and fat plasma rings found in
\cite{Lahiri:2007ae}. The energy is fixed at $\tE=40$ as in
\cite{Lahiri:2007ae}. \label{fig:phasediagram}}
\end{figure}

We have integrated numerically equation (\ref{YL:profiles}) to find
lobed-like solutions. In the numerical search, we fix the energy at
$\tilde{E}=40\pm 10^{-7}$ (for simplicity, since it allows
comparison with previous investigations
\cite{Lahiri:2007ae,Bhardwaj:2008if,Cardoso:2009bv} ), and vary
$\tO$. There are three parameters to be varied in the numerical
integration: $\tO,\,k$ and $\vo(0)$. For each $\tO$, we search for
an array of $k,\vo(0)$, which yield $m$-lobed solutions with
$\vo(0)/\vo(2\pi)=1\pm 10^{-7}$. Among those, we select the one for
which $\tilde{E}=40\pm 10^{-7}$. We started the integration close to
the plasma ball configuration, by looking for a two-lobed solution
with $\vo(0)\sim \vo(\pi)$. We find that the bifurcation to plasma
peanuts occurs at the critical values $\tJ\simeq 22.2$, $\tO\simeq
0.16$, $\tS\simeq 32.7$, and $\tilde{T}\simeq 1.0$. Our results are
summarized in Figures \ref{fig:shape}-\ref{fig:phasediagram}.

In Figure \ref{fig:shape}, we show typical two-lobed configurations
for different values of rotation parameter, $\tO\sim 0.13,0.12,0.09$
respectively. As $\tO$ decreases, the configurations get more and
more distorted,  and for small rotations they assume the form of
peanut-like shapes, as in the rightmost panel of Figure
\ref{fig:shape}.
Once one knows the configuration shape it is straightforward to
compute the energy, entropy and angular momentum from eqs.
(\ref{ChargesPeanutPert}) and to establish the phase diagrams
$\tS(\tJ)$, $\tO(\tJ)$ and $\tilde{T}(\tJ)$. These diagrams are
shown in Figure \ref{fig:phasediagram}, where we include also plasma
balls and plasma rings \cite{Lahiri:2007ae}. For the particular
energy we are considering, the 2-lobed configurations have a slope
of about $\p\tS/\p\tJ\sim -0.149$. This is in very good agreement
($\sim 5\%$) with the perturbative construction prediction
$\p\tS/\p\tJ \simeq -0.142$ of the last subsection. We have
constructed numerically other configurations, including classical
configurations, for which the velocities are very small, and we have
obtained good agreement for the slopes predicted in
\cite{Cardoso:2009bv}.

One particularity of the plasma peanuts (common to the thin rings) may be noted here:
for these configurations an increase of angular momentum produces a {\em decrease} of
angular velocity, as is evident from Figure~\ref{fig:phasediagram}. Such a phenomenon
is made possible because the moment of inertia grows accordingly and this is apparent
in Figure~\ref{fig:shape}.

%%%%%%%%%%%%%%%%%%%%%%%%%%%%%%%%%%%%%%%%%%%%%%%%%%%%%%%%%%%%%%%%%%%%%%%%%%%%%%%%%%%%%
\subsection{No $m$-lobed plasma rings \label{sec:NoLobedRings}}
%%%%%%%%%%%%%%%%%%%%%%%%%%%%%%%%%%%%%%%%%%%%%%%%%%%%%%%%%%%%%%%%%%%%%%%%%%%%%%%%%%%%%

A natural question that emerges from the plasma ball analysis done
so far is whether or not it is possible to have also $m$-lobed plasma
rings. In a previous study this question was addressed by perturbing
hydrodynamic equations~\eqref{continuity:diss}-\eqref{YoungLap:diss}
with an ansatz that satisfied the symmetries of the hypothetical $m$-lobed plasma
rings \cite{Cardoso:2009bv}. The analysis concluded that plasma rings are stable against
this particular sector of perturbations. We now revisit this problem
from a different perspective. The outcome of the present analysis will reinforce the
conclusion of the previous study.

Motivated by the success of the analysis done in subsection
\ref{sec:peanutsPert}, the idea is to ask if we can get $m$-lobed
plasma rings as perturbative solutions around a possible bifurcation
point in the plasma ring branch.
It follows from the Young-Laplace equation \eqref{YoungLap:Equil}
and equilibrium equation of state \eqref{DpEquil} that the outer
boundary of a hypothetical $m$-lobed plasma ring has to satisfy
\eqref{YL:profiles}, and its inner boundary must obey a similar
relation with the trade $\vo\rightarrow \vi$ for fixed $\{ k, \tO
\}$. The fact that the parameter $k=\rho_*/(3\rho_0)$ must be the
same for both boundaries in particular implies that such a solution
would have to satisfy the relation,
\begin{eqnarray}\label{YL:LobedRings}
k&=&(1-\vo^2)^2 \lp 1-\tO\,\frac{\vo \vo''
(1-\vo^2)-\vo'^{\,2}(2-\vo^2)-\vo^2}{\lpp \vo^2+\vo'^{\,2}(1-\vo^2)
\rpp^{3/2}} \rp \nonumber\\
&=&
 (1-\vi^2)^2 \lp 1-\tO\,\frac{\vi \vi''
(1-\vi^2)-\vi'^{\,2}(2-\vi^2)-\vi^2}{\lpp \vi^2+\vi'^{\,2}(1-\vi^2)
\rpp^{3/2}} \rp\,.
\end{eqnarray}
In the axisymmetric case, $\vo'=0=\vi'$, this relation reduces to
the equilibrium condition \eqref{Ring:gConstraint}.

Consider now the (natural) possibility that an $m$-lobed plasma ring
could emerge from a slight perturbation of an axisymmetric plasma
ring. Following a similar strategy as in section
\ref{sec:peanutsPert}, we linearize \eqref{YL:LobedRings} by
perturbing the boundaries as
\begin{equation} v_\alpha(\psi) = \widehat{v}_\alpha \lpp 1+\varepsilon\, \nu_\alpha(\psi) +
\mathcal{O}(\varepsilon^2)\rpp \,, \quad {\rm with} \quad
\alpha={\rm o,i}\,,
\end{equation}
 and $\tO$ and $k$ as in \eqref{Ball:Perturb}.
Again, the hat refers to the (constant) unperturbed quantities. To
leading order, we get \eqref{Ring:gConstraint} and in the
next-to-leading order we find the two equations,
 \be \label{perturb:eigenRing} \nu_\alpha'' + m_\alpha^2 \nu_\alpha = \Delta_\alpha
\,, \quad {\rm with} \quad \alpha={\rm o,i}\,, \ee
 and
\begin{equation} \label{Ring:m} m_\alpha^2 = \frac{1}{1-v_\alpha^2} + \epsilon_\alpha \frac{4\,\widehat{k} \,\widehat{v}_\alpha^3}{\widehat{\Omega}
\lp 1-\widehat{v}_\alpha^2\rp^4} \,, \qquad \Delta_\alpha=\epsilon_\alpha
\widehat{v}_\alpha\,
 \frac{\lpp \widehat{k}-\lp 1-\widehat{v}_\alpha^2\rp^2 \rpp \omega-\widehat{\Omega} \, \kappa}{\widehat{\Omega}^2\lp 1-\widehat{v}_\alpha^2\rp^3 }
 \,,\qquad \epsilon_{\rm o,i}\equiv\pm 1 \,.
\end{equation}

\begin{figure}[ht]
\begin{tabular}{ccc}
\epsfig{file=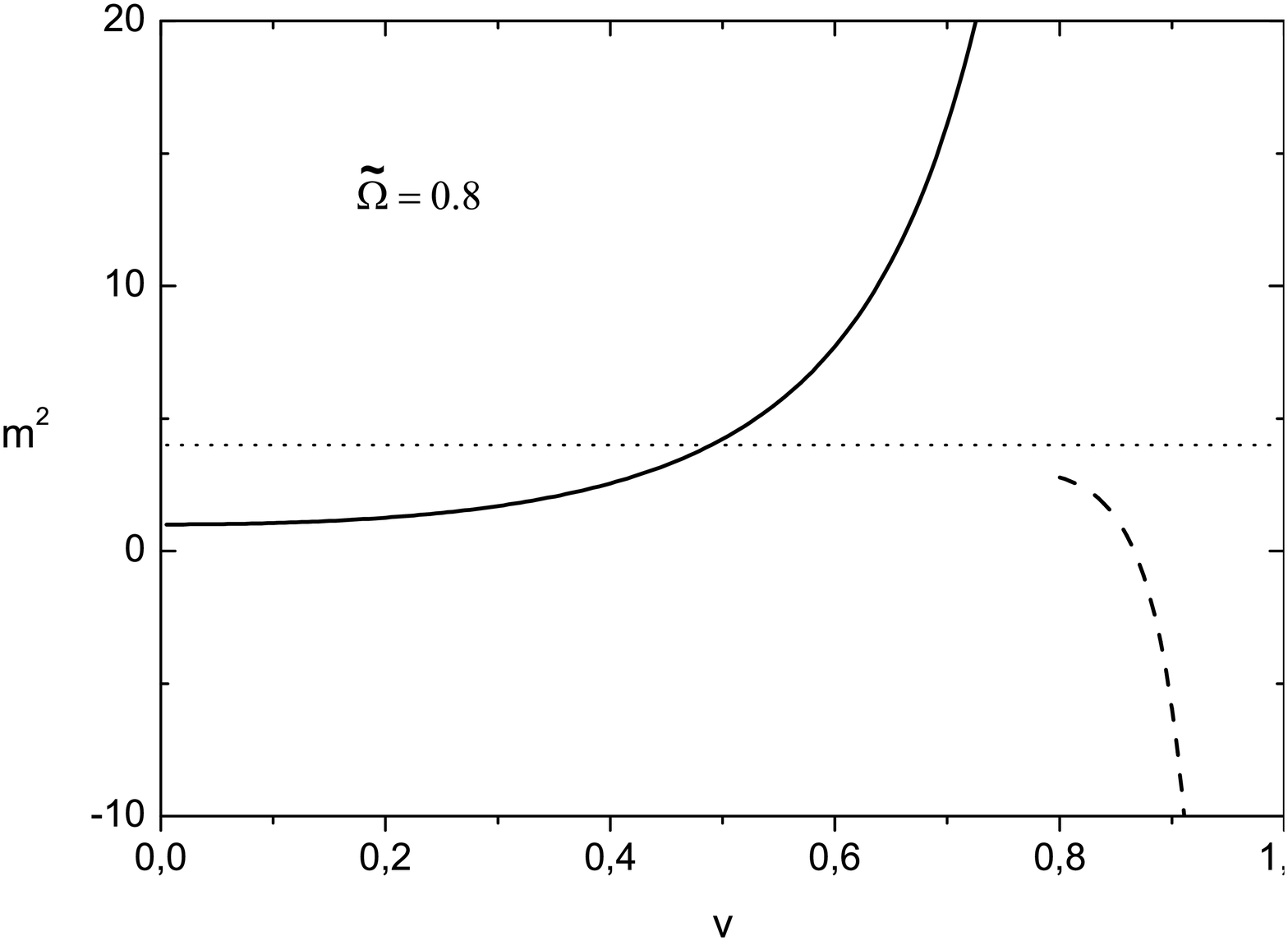,width=5cm,angle=0}&\epsfig{file=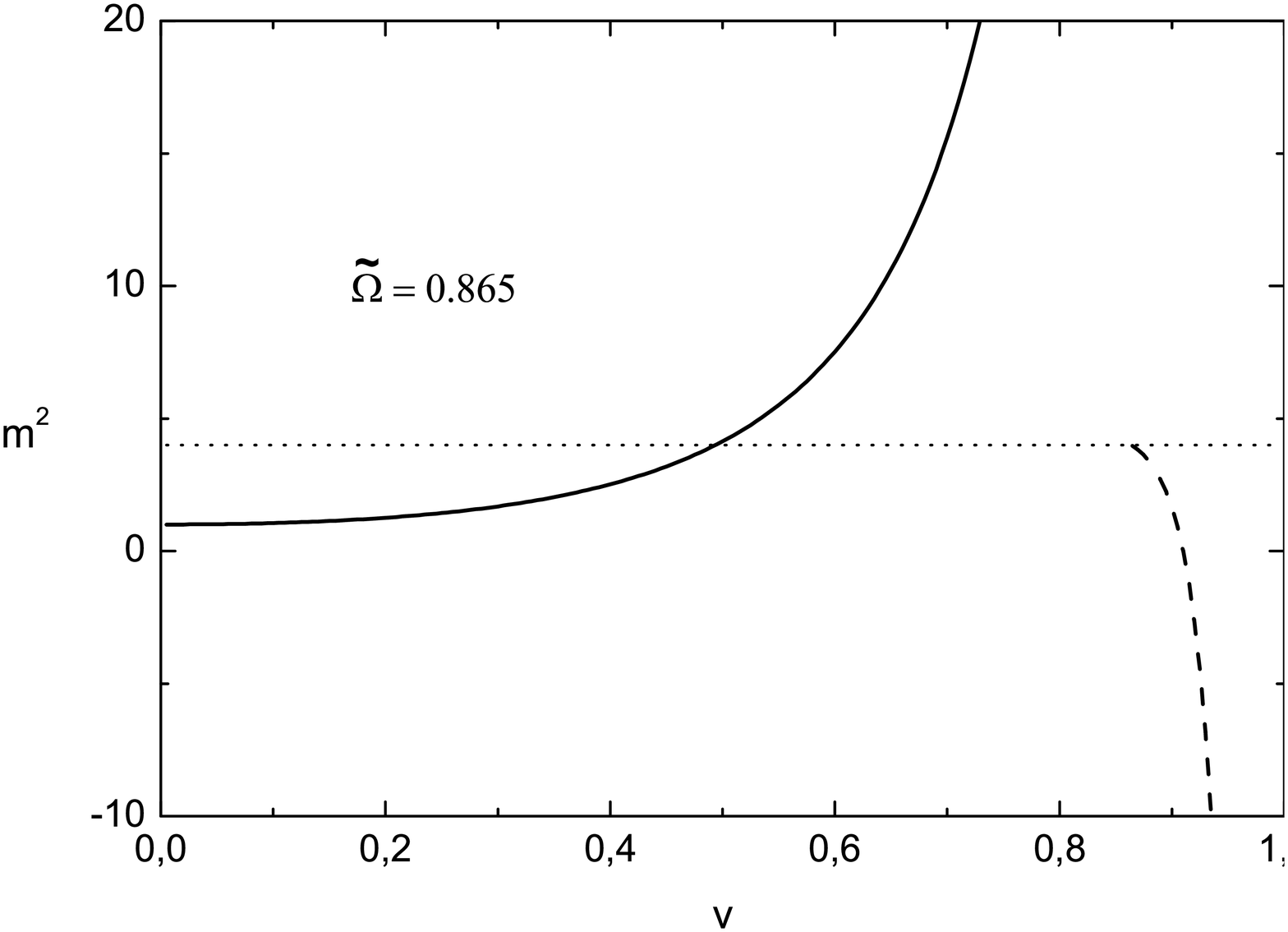,width=5cm,angle=0}&\epsfig{file=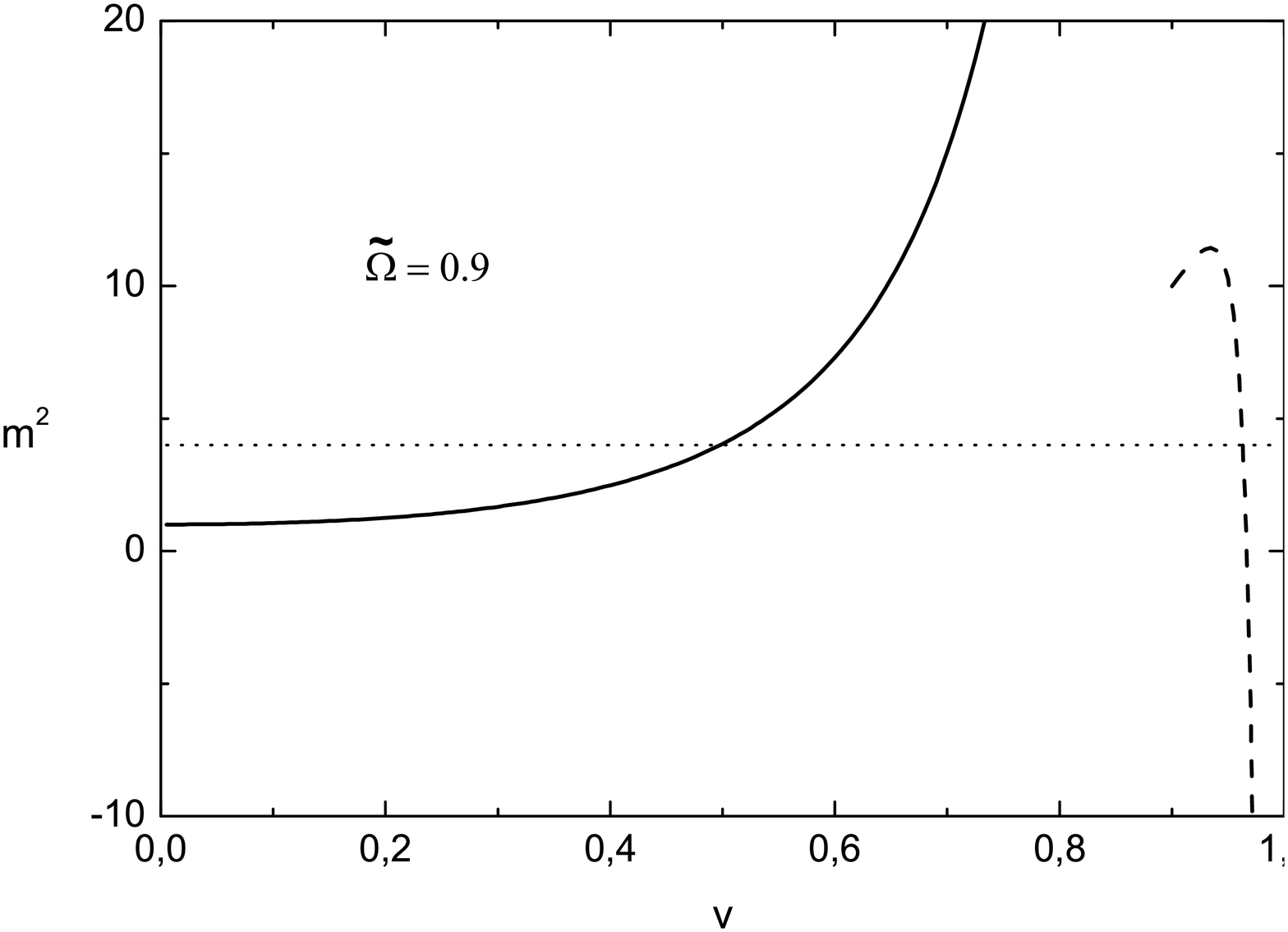,width=5cm,angle=0}
\end{tabular}
\caption{$m^2$ as a function of $v_\alpha$ for the three
possible regimes of $\widehat{\Omega}$. The solid line represents
$m_{\rm o}^2(\widehat{\Omega},\widehat{v}_\mathrm{o})$, while the dashed line represents
$m_{\rm i}^2(\widehat{\Omega},\widehat{v}_\mathrm{i})$. The case  $m_{\rm o}= m_{\rm i}=2$ is possible
only for large rotations and when $\widehat{v}_\mathrm{i} \geq \widehat{v}_\mathrm{o}$. This violates the
plasma ring condition that the inner radius must be smaller than the
outer radius. \label{fig:PeanutRing}}
\end{figure}

Each of these equations admits an oscillatory solution of the form
\eqref{OscSolution}.  $m$-lobed plasma rings can exist only if the
number of nodes in the outer and inner boundaries is the same, \ie
if the condition
\begin{equation} \label{RingPeanutCond}
 m_{\rm o}= m_{\rm i} \,,\qquad m_{\rm o}, m_{\rm i} \in  \mathbb{Z}
\end{equation}
is satisfied. In this relation and \eqref{Ring:m}, the unperturbed
quantities are constrained as discussed in subsection
\ref{sec:ballsRings}, \eg one must have $\widehat{\Omega} \leq \widehat{v}_\mathrm{i}\leq 1$ and
$\widehat{k}=\widehat{k}(\widehat{\Omega},\widehat{v}_\mathrm{o})=\widehat{k}(\widehat{\Omega},\widehat{v}_\mathrm{i})$ as indicated
in \eqref{Ring:gConstraint}. An inspection of \eqref{Ring:m}
concludes that the equality \eqref{RingPeanutCond} is verified only
when $\widehat{v}_\mathrm{i}\geq \widehat{v}_\mathrm{o}$ (see Fig. \ref{fig:PeanutRing}). This clearly
violates the basic assumption that the inner radius of the plasma
ring must be smaller than its outer radius.

%\begin{figure}[t]
%\centerline{\includegraphics[width=.30\textwidth]{PeanutRing1.pdf}
%\hfill\includegraphics[width=.30\textwidth]{PeanutRing2.pdf}
%\hfill\includegraphics[width=.30\textwidth]{PeanutRing3.pdf} }
%\caption{\small $m^2$ as a function of $v$ and for the three
%possible regimes of $\Omega$. The solid line represents
%$m_o^2(\Omega,\vo)$, while the dashed line represents
%$m_i^2(\Omega,\vi)$. The case  $m_{\rm o}= m_{\rm i}=2$  is possible
%only for  large rotations and when $\vi\geq \vo$. This violates the
%plasma ring condition that the inner radius must be smaller than the
%outer radius.} \label{fig:PeanutRing}
%\end{figure}

%
%
%
%\begin{figure}[ht]
%\begin{tabular}{c}
%\epsfig{file=Plots/RingMerged.eps,width=10cm,angle=0}
%\end{tabular}
%\caption{$m^2$ as a function of $v$ and for the three
%possible regimes of $\Omega$. The solid line represents
%$m_o^2(\Omega,\vo)$, while the dashed line represents
%$m_i^2(\Omega,\vi)$. The case  $m_{\rm o}= m_{\rm i}=2$  is possible
%only for  large rotations and when $\vi\geq \vo$. This violates the
%plasma ring condition that the inner radius must be smaller than the
%outer radius. \label{fig:PeanutRing}}
%\end{figure}
%
This analysis therefore rules out the possibility that lobed plasma
rings could exist as a branch of solutions emerging perturbatively
from the axisymmetric plasma rings. There is a clear physical reason
for why the plasma rings are stable against ``lobing'', in contradiction
with plasma balls. The additional angular momentum acts to push
the fluid radially outward. For the plasma ball the only way this can
be done while preserving the energy is by breaking axial symmetry, whereas
plasma rings can accommodate such a change by becoming larger and
thinner\footnote{This argument is due to V. Hubeny, to whom we thank.}.

Reference \cite{Cardoso:2009bv} looked for the
possible existence of these solutions following a different path. It
analyzed the stability of a plasma ring against $m$-lobed
perturbations such that these were in phase both in the inner and
outer boundary deformations. No unstable mode was found. These two
negative results following different approaches seem to suggest that
the existence of $m$-lobed plasma rings is ruled out.

%%%%%%%%%%%%%%%%%%%%%%%%%%%%%%%%%%%%%%%%%%%%%%%%%%%%%%%%%%%%%%%%%%%%%%%%%%%%%%%%%%%%%
\subsection{The hydrodynamic regime \label{sec:HydroRegime}}
%%%%%%%%%%%%%%%%%%%%%%%%%%%%%%%%%%%%%%%%%%%%%%%%%%%%%%%%%%%%%%%%%%%%%%%%%%%%%%%%%%%%%

Relativistic hydrodynamics provides a good effective description of
the deconfined plasma phase of ${\cal N}=4$ Yang Mills theory
compactified down to $d=3$ on a Scherk-Schwarz circle only if
certain conditions are satisfied \cite{Lahiri:2007ae}. First,
hydrodynamics is by definition valid when the thermodynamic
quantities of the fluid vary on a lengthscale large when compared
with the mean free path, $\ell_{\rm mfp}$, of fluid constituent
particles. In our case $\ell_{\rm mfp}\sim  T_c^{-1}  \sim
\frac{\sigma}{\rho_0}$. A good estimate for the valid regime is
obtained when the maximum fractional rate of change of the fluid
local temperature, $\frac{\delta \tloc}{\tloc}{\bigl |}_{\rm
max}\sim \partial_r \ln\gamma {\bigl |}_{\rm max}$ (recall that
$\tloc=T\gamma$) is much smaller than $\ell_{\rm mfp}^{-1}$. This occurs
for $\frac{\tO \vo}{1-\vo^2} \ll 1$. This condition is satisfied by
a wide range of plasma balls and rings as long as we are away from
extremality \cite{Lahiri:2007ae}. Second, the analysis done so far
and onwards assumes that surface tension is constant,
$\sigma=\sigma( T_c )$, when in fact it is a function of the fluid
temperature at the surface. This assumption is valid when $\tloc/
T_c \sim 1$ at the boundary surfaces. This is the case for an extended
range of energies and angular momenta as long as we do not approach
the extremal configurations \cite{Lahiri:2007ae} too closely. Finally,
the boundary of the plasma is treated as a delta-like surface when
in fact it has a thickness of order $ T_c ^{-1}$. So the analysis is
valid when the boundary radius is everywhere large when compared
with $ T_c ^{-1}$, $\{\ro,\ri,\ro-\ri\} \gg \frac{\sigma}{\rho_0}$.
This is satisfied if the plasma energy is large and again if we are
away from the extremal configurations \cite{Lahiri:2007ae}.

%%%%%%%%%%%%%%%%%%%%%%%%%%%%%%%%%%%%%%%%%%%%%%%%%%%%%%%%%%%%%%%%%%%%%%%%%%%%%%%%%%%%%
\section{Gravity duals. Discussion \label{sec:Discussion}}
%%%%%%%%%%%%%%%%%%%%%%%%%%%%%%%%%%%%%%%%%%%%%%%%%%%%%%%%%%%%%%%%%%%%%%%%%%%%%%%%%%%%%

Since plasma balls are unstable above a critical rotation rate, the
holographic dual SS AdS$_5$ black holes should also be unstable
against $m$-lobed perturbations. This instability then provides a
mechanism that dynamically bounds the rotation of these black holes.
From the plasma results, we expect this bound to be well below the
rotation where the extreme solution is reached (for which $\widetilde{T}=0$), and below the
minimal rotation where SS AdS$_5$ black rings can exist. Very little
is known about SS AdS$_5$ black holes, so we are unable at this
point to explicitly check the existence of this instability and
rotation bound in the gravitational system.

In the fluid, we explicitly found the new branch of stationary
non-axisymmetric plasmas bifurcating from the plasma ball curve
at the point where the instability becomes active. In the simplest
$m=2$ case this is a plasma peanut. In the entropy {\it vs} angular
momentum diagram (for fixed mass) the plasma peanuts have higher
entropy than the plasma ball. In fact, for the range of angular
momenta for which plasma peanut solutions exist, they are
entropically dominant over all the stationary configurations.

In the dual gravitational system, $m$-lobed plasma balls are in
correspondence with $m$-lobed black holes. However, the latter can
only exist as a long lived object, but never as a stationary
solution. Indeed, a rotating non-axisymmetric black hole has a
quadrupole moment and thus necessarily radiates away its lobed
deformations\footnote{A static non-axisymmetric black hole was
argued to exist in Kaluza-Klein gravity, \ie that asymptotes to
$M^d\times S^1$, for $d\geq 4$ \cite{Dias:2007hg}.}. The expectation
is that this emission of gravitational waves will proceed until the
system is back to an axisymmetric configuration spinning at a rate
below the critical rotation where the instability kicks in.

This point deserves a few more words. Consider first global
AdS$_{d+2}$ spacetime. Its boundary is the Einstein static universe
$\mathbb{R}_t\times S^{d}$. This boundary behaves effectively as a
reflecting wall or box. Therefore, in this background one could
eventually have a spinning non-axisymmetric black hole surrounded by
rotating radiation. The rate at which the black hole would be
radiating could in principle be balanced by the absorption of
previously emitted radiation that is reflected from the boundary. An
equilibrium solution with a black hole plus orbiting radiation is
{\it a priori} conceivable in this background. On the other hand, SS
AdS has boundary $\mathbb{R}_t\times \mathbb{R}^{d-1}\times
S_{SS}^{1}$ and behaves quite differently. In this background, only
propagation along the holographic radial direction faces reflective
boundary conditions. Along the other spacelike directions, namely
those parallel to the holographic boundary (where the dual fluid
lives), we have asymptotically flat boundary conditions. It then
follows that waves emitted by a black hole along the radial
direction will be reflected back but radiation emitted along the
boundary directions will leak towards infinity. Therefore, the
$m$-lobed plasma balls are dual to long-lived lobed SS AdS black
holes that decay slowly, \ie that loose angular momentum along the
boundary directions and cannot be stationary.

The existence of the lobed instability and of the stationary plasma
peanuts and related families exposes the need to explore one point
in the fluid/gravity correspondence that is still poorly understood:
what is the fluid description of gravitational interactions and
gravitational radiation? At the approximation level at which the duality
is currently understood, hydrodynamics can only describe
gravitational systems where these phenomena are suppressed. Indeed,
two plasma balls do not interact, their collision is not
accompanied by radiation emission and a non-axisymmetric plasma
ball does not radiate. In short, the fluid description only captures
the interaction of two black holes in the limit when their separation
is much larger than the AdS radius, and the gravitational emission
in the limit where its wavelength is much larger than the AdS scale.

It is known that the gravitational radiation and interaction
corresponds in the plasma to the emission of glueballs. It is also
known that in the large $N$ limit of confining $SU(N)$ gauge
theories that have a gravitational dual, the energy loss in this
emission process is suppressed by a $1/N^2$ factor when compared
with the plasma ball energy density. For this reason we know that
our plasma results describe a necessarily long-lived SS AdS lobed
black hole where gravitational emission occurs at a very slow rate.
It would nevertheless be very interesting to work out more quantitatively
the fluid description of the gravitational emission and interaction
processes, to accommodate also in the formulation cases where these
phenomena are not suppressed. However, it is conceivable that such
processes might force one to leave the regime of validity of the
hydrodynamic description and therefore may be hard to capture\footnote{We
thank M. Rangamani for pointing this out to us.}.

Although we have only worked out the lowest dimensional case, higher
dimensional SS AdS theories also have a fluid description. More
specifically, for any $d\geq 3$, the SS compactification of
$(d+1)$-dimensional CFT has a $d$-dimensional fluid dynamic
description which is dual to a SS compactification of AdS$_{d+2}$.
An interesting evolution is expected as we climb the dimension
ladder. As described above, the asymptotics of SS AdS gravity
roughly interpolates between AdS and Minkowski asymptotics depending
on the spatial direction we look at. But as we increase the number
of dimensions the ``AdS-like" radial direction is kept and we are
only adding more and more ``flat-like" boundary directions.
Therefore, we antecipate that as $d$ grows the fluid results should
describe black holes whose properties increasingly resemble those of
asymptotically flat black holes
\cite{Caldarelli:2008mv,Bhattacharya:2009gm}. The known results
confirm this expectation, as we briefly review next. Start with
$d=3$. The phase diagram for plasma balls and plasma rings in $3d$
(see Figure \ref{fig:phasediagram}) is similar to the phase diagram
expected for black holes and black rings in AdS$_5$
\cite{Lahiri:2007ae}. In particular, in contrast to the
asymptotically flat solutions, the black rings have an upper bound
on the rotation above which they would not fit in the AdS
box~\cite{Lahiri:2007ae}. So, the $3d$ plasma results indicate that
the SS AdS$_5$ black objects behave similarly to AdS$_5$ black
holes. When we increase the dimension by one unit two interesting
new features emerge \cite{Lahiri:2007ae,Bhardwaj:2008if}. (The
following discussion only highlights some properties of interest for
our purpose; we ask the reader to see \cite{Bhardwaj:2008if} for
details). First, the plasma rings have now unbounded angular
momentum. That is, SS AdS$_6$ black rings behave much like $6d$
asymptotically flat black rings. So, quite amazingly, in what
concerns this particular feature in the fluids $d=4$ sets already
the critical dimension for the above mentioned transition between an
asymptotically AdS and asymptotically flat behavior. Second, from
the gravitational perspective, we know that asymptotically flat
(singly spinning) black holes in $d\geq 6$ have no bound on their
rotation. The transition from moderately rotating black holes into
ultraspinning ones is set by the critical rotation where the
so-called ultraspinning instability becomes active
\cite{Emparan:2003sy,Dias:2009iu}\footnote{In more detail, when
rotation keeps increasing in the ultraspinning regime the horizon
flattens in the neighborhood of the axis poles and the geometry
becomes more and more like the geometry of a black membrane
\cite{Emparan:2003sy}. But the latter is unstable against the
Gregory-Laflamme instability when their length along the extended
directions is bigger than their transverse radius
\cite{Gregory:1993vy}. Therefore, at the critical rotation where we
enter the ultraspinning regime (and well below the limit where the
geometry becomes a black membrane) the rotating black holes are
expected to become unstable against the naturally dubbed
ultraspinning instability \cite{Emparan:2003sy}. A recent study
confirmed that this is indeed the case \cite{Dias:2009iu}. In the
fluid description, the Gregory-Laflamme instability maps into the
Rayleigh-Plateau instability that is responsible for the pinch-off
of long plasma tubes or branes into plasma balls
\cite{Cardoso:2006ks,Caldarelli:2008mv}.}. In the phase diagram of
solutions this marks a bifurcation point to a new branch of
axisymmetric pinched black holes whose horizon is distorted with
ripples along the polar direction. Returning to the fluid results,
in $d=4$  a third stationary solution $-$ the pinched plasma balls
$-$ is indeed found, in addition to the plasma balls and rings
\cite{Lahiri:2007ae,Bhardwaj:2008if}.

We have not worked out the technical details but in $d>3$ the
$m$-lobed instability must also be present in the plasma balls. The
reason being that it is well established that unstable phenomena
that are already present in classical fluid dynamics cannot cease to
exist in relativistic hydrodynamics. Now, in \cite{Cardoso:2006sj}
it was found that the lobed instability is indeed active in higher
dimensional classical fluids. Therefore it should persist in the
relativistic regime. The interesting question is then whether this
instability appears at lower or higher values of rotation than the
ultraspinning one. We argue that it should appear for lower spins.
The argument is simple but seems to be robust: in $3d$ the former
instability is present while the latter does not appear. It would
become active at rotations higher than the upper bound for the
angular momentum. When we increase $d$ this cap in $J$ disappears
and the ultraspinning makes its appearance but, it should do so at a
rotation larger than the $2$-lobed critical rotation.

As emphasized several times, the plasma results are strictly dual to
SS AdS gravity. But as described above, in certain regimes, we also
have or expect strong similarities between black holes in this
background and those that are globally AdS or asymptotically flat.
It is therefore compelling to conjecture the possibility that a
non-axisymmetric $m$-lobed instability might also be present in
higher dimensional Myers-Perry($-$AdS) black holes. We do not find
any reason to discard it {\it a priori}; a numerical investigation
of this hypothetic instability might reveal interesting results.

%%%%%%%%%%%%%%%%%%%%%%%%%%%%%%%%%%%%%%%%%%%%%%%%%%%%%%%%%%%%%%%%%%%%%%%%%%%%%%%%%%%%%%%%%%%%%%%%%%%%%%%%%%%%%%%%%
\section*{Acknowledgments}

We warmly thank Marco Caldarelli and R. Loganayagam for very fruitful
discussions, and specially Veronika Hubeny and Mukund Rangamani for
their useful comments to the final version of this manuscript. OJCD
thanks the organizers and participants of the {\it Workshop on
Fluid-Gravity Correspondence}, University Ludwig-Maximilians of
Munich, Germany, for wonderful hospitality and discussions. OJCD
acknowledges financial support provided by the European Community
through the Intra-European Marie Curie contract PIEF-GA-2008-220197.
JVR acknowledges financial support from {\it Funda\c{c}\~ao para a
Ci\^encia e Tecnologia} (FCT)-Portugal through fellowship
SFRH/BPD/47332/2008. This work was partially funded by FCT-Portugal
through projects PTDC/FIS/64175/2006, PTDC/ FIS/098025/2008,
PTDC/FIS/098032/2008, CERN/FP/ 83508/ 2008. The authors thankfully
acknowledge the computer resources, technical expertise and
assistance provided by the Barcelona Supercomputing Center - Centro
Nacional de Supercomputación.

%%%%%%%%%%%%%%%%%%%%%%%%%%%%%%%%%%%%%%%%%%%%%%%%%%%%%%%%%%%%%%%%%%%%%%%%%%%%%%%%%%%%%%%%%%%%%%%%%%%%%%%%%%%%%%%%%
\appendix

%\section*{Appendices}
\section*{Appendix}
%%%%%%%%%%%%%%%%%%%%%%%%%%%%%%%%%%%%%%%%%%%%%%%%%%%%%%%%%%%%%%%%%%%%%%%%%%%%%%%%%%%%%%%%%%%%%%%%%%%%%%%%%%%%%%%%%

%%%%%%%%%%%%%%%%%%%%%%%%%%%%%%%%%%%%%%%%%%%%%%%%%%%%%%%%%%%%%%%%%%%%%%%%%%%%%
\setcounter{equation}{0}
\section{Non-relativistic lobed configurations \label{sec:NRpeanuts}}
%%%%%%%%%%%%%%%%%%%%%%%%%%%%%%%%%%%%%%%%%%%%%%%%%%%%%%%%%%%%%%%%%%%%%%%%%%%%%

In this Appendix we review the classical hydrodynamic equations that
describe non-relativistic fluids and, in particular, lobed
configurations \cite{chandra65}-\cite{hilleaves}. As a consistency
check, we confirm that the non-relativistic (NR) limit of our
equations for the plasma peanuts reduces to these classical
equations.

The classical analysis of rigidly rotating fluids is standardly done
in the rotating frame of reference. The advantage of this frame is
that the (unperturbed) velocity of the fluid vanishes.  In this
frame, the classical Navier-Stokes, continuity and Young-Laplace
equations are respectively given by
\begin{eqnarray}
&& \partial_t {\bf v} +({\bf v}\cdot\overline{\nabla}) {\bf v}=  -
\frac{1}{\rho} \overline{\nabla} P -2{\bf \Omega}\times {\bf v}
+\frac{1}{2}\overline{\nabla}\lp |{\bf \Omega}\times {\bf r}|\rp\,,
\nonumber\\
&& \partial_t\rho +{\bf v} \cdot \overline{\bf \nabla} \rho +\rho
\overline{\nabla} \cdot {\bf v}=0\,,
\nonumber\\
&&    P_<-P_> =\sigma \overline{K}\,, \qquad {\rm with}\quad
\overline{K}\equiv \overline{\nabla}\cdot {\bf n}\,,
 \label{NR:HydroEqs}
 \end{eqnarray}
where  ${\bf v}$ is the spatial fluid velocity, and
$\overline{\nabla}$ the spatial gradient. In the Navier-Stokes
equation, the two last terms describe, respectively, the Coriolis
and the centrifugal acceleration contributions. We are interested in
stationary fluids so we did not include dissipation terms that
vanish for these configurations. We consider incompressible fluids
(\ie with vanishing convective derivative, $d\rho/dt\equiv
\partial_t\rho +{\bf v} \cdot \overline{\bf \nabla} \rho=0$) and thus
the continuity equation reduces to the statement that the velocity
is a solenoid vector, $\overline{\nabla} \cdot {\bf v}=0$. Note that
in the classical Young-Laplace equation, $\overline{K}$ is the
spatial or mean curvature (and not the spacetime curvature).

For a fluid in rigid rotation, the Navier-Stokes equation yields for the
pressure: $P=\frac{1}{2}\rho \Omega^2 r^2 +C$, where $C$ is a
constant. To make later contact with the relativistic result, we
choose to replace the constant quantities $\{\rho,C \}$ by
$\{\rho_0,k \}$, with the relation between them being
$C=\rho_0(k-1)$ and $\rho=4\rho_0 k$. A stationary fluid with a
non-axisymmetric boundary profile described by
$f(r,\phi)=r-R(\phi)=0$ has mean curvature
$\overline{K}=\frac{R^2+2R'^{\,2}-R R''}{\lp R^2+R'^{\,2}
\rp^{3/2}}$. Introducing the dimensionless quantities
\begin{equation}\label{NR:newvars}
   \psi=\frac{\phi}{\Omega}\,,\qquad \vo(\psi) = \Omega R(\psi)\,,
    \qquad \tO = \frac{\sigma\Omega}{\rho_0} \,,
\end{equation}
the Young-Laplace equation can then be written as
\begin{equation}\label{NR:YLprofiles}
  \frac{\vo \vo'' -2\vo'^{\,2}-\vo^2}{\lp
\vo^2+\vo'^{\,2} \rp^{3/2}}
  +\frac{1}{\tO}\lpp (k-1)+2k \vo^2 \rpp =0\,.
\end{equation}
This equation agrees with the NR limit of \eqref{YL:profiles},
as it should, and gives the equation for the profile of stationary
classical lobed configurations. It has the first integral,
\be \label{NR:1stIntegral}
 Q_{\rm NR}\equiv \frac{\vo^2}{
\sqrt{\vo^2+\vo'^{\,2}} }
  -\frac{1}{2\tO}\lpp k(1+\vo^2)^{2}-(k+1)\vo^2\rpp,
\ee
obtained by integration of \eqref{NR:YLprofiles}. By curiosity, note
that this is an example of a case where ``the limit of an expression is
not the expression of the limit'' since \eqref{NR:1stIntegral} is
{\it not} the NR limit of the first integral \eqref{1stIntegral}.

%%%%%%%%%%%%%%%%%%%%%%%%%%%%%%%%%%%%%%%%%%%%%%%%%%%%%%%%%%%%%%%%%%%%%%%%%%%%%%%

%\begin{thebibliography}{99}

\providecommand{\href}[2]{#2}\begingroup\raggedright\endgroup

%\end{thebibliography}

\end{document}